\documentclass[aps,pra,twocolumn,footinbib,superscriptaddress]{revtex4-1} %showpacs
	
	\usepackage{textcomp}
	\usepackage{amsmath}
	\usepackage{amssymb}
	\usepackage{graphicx} 
	\usepackage[breaklinks=true,colorlinks,citecolor=blue,linkcolor=blue,urlcolor=blue]{hyperref}
	\usepackage{mathtools}
	\usepackage{bm}
	\usepackage{color}
	\usepackage{ulem}
	\usepackage{braket}
	\usepackage{lipsum}
    \usepackage{lmodern}
    \usepackage[T1]{fontenc}
	\usepackage{changes}
	
\newcommand{\bg}{\begin{equation} \begin{gathered}}
\newcommand{\eg}{\end{gathered} \end{equation}}
\newcommand{\ba}{\begin{equation}}
\newcommand{\ea}{\end{equation}}
	
	\newcommand{\e}{e}
	\renewcommand{\i}{i}

	\newcommand{\Tr}{\mathop{\text{Tr}}\nolimits}

	\newcommand{\red}[1]{{\color{red} #1}}

\begin{document}
\title{Maximum power heat engines and refrigerators in the fast-driving regime}

\author{Vasco Cavina} \email{vasco.cavina@uni.lu}
\affiliation{NEST, Scuola Normale Superiore and Istituto Nanoscienze-CNR, I-56126 Pisa, Italy}
\affiliation{Complex Systems and Statistical Mechanics, Physics and Materials Science,University of Luxembourg, L-1511 Luxembourg, Luxembourg}

\author{Paolo A. Erdman}
\affiliation{NEST, Scuola Normale Superiore and Istituto Nanoscienze-CNR, I-56126 Pisa, Italy}
\affiliation{Freie Universit{\" a}t Berlin, Department of Mathematics and Computer Science, Arnimallee 6, 14195 Berlin, Germany}

\author{Paolo Abiuso}
\affiliation{ICFO – Institut de Ci{\`e}ncies Fot{\`o}niques, The Barcelona Institute of Science and Technology, 08860 Castelldefels (Barcelona), Spain}

\author{Leonardo Tolomeo}
\affiliation{
Mathematical Institute, Hausdorff Center for Mathematics, Universit{\"a}t Bonn, 456-321 Bonn, Germany}

\author{Vittorio Giovannetti}
\affiliation{NEST, Scuola Normale Superiore and Istituto Nanoscienze-CNR, I-56126 Pisa, Italy}

\begin{abstract}

We study the optimization of the performance of arbitrary periodically driven thermal machines. Within the assumption of fast modulation of the driving parameters, we derive the optimal cycle that universally maximizes the extracted power of heat engines, the cooling power of refrigerators, and in general any linear combination of the heat currents. We denote this optimal solution as ``generalized Otto cycle'' since it shares the basic structure with the standard Otto cycle, but it is characterized by a greater number of fast strokes. We bound this number in terms of the dimension of the Hilbert space of the system used as working fluid.
The generality of these results allows for a widespread range of applications, such as reducing the computational
complexity for numerical approaches, or obtaining the explicit form of the optimal protocols when the system-baths interactions are characterized by a single thermalization scale.
In this case, we compare the thermodynamic performance of a collection of optimally driven non-interacting and interacting qubits. Remarkably, for refrigerators the non-interacting qubits perform almost as well as the interacting ones, while in the heat engine case there is a many-body advantage both in the maximum power, and in the efficiency at maximum power. 
Additionally, we illustrate our general results studying the paradigmatic model of a qutrit-based heat engine. Our results strictly hold in the semiclassical case in which no coherence is generated by the driving, and finally we discuss the non-commuting case.
\end{abstract}


%72.20.Pa 	Thermoelectric and thermomagnetic effects
%73.23.-b		Electronic transport in mesoscopic systems

\maketitle

\section{Introduction}
The most important thermal machines that can be constructed utilizing two or more thermal baths are the heat engine and the refrigerator. These machines are mainly characterized by two figures of merit: the efficiency (or coefficient of performance for the refrigerator) and the extracted power (or cooling power). The optimal strategy to maximize the efficiency (and the coefficient of performance) was identified already in the $19^\text{th}$ century, and it is closely related to the second law of thermodynamics. As such it is characterized by a universal strategy: \textit{infinitely slow} transformations, known as reversible transformations, must be performed \cite{Huang1987}. On the other hand, the maximization of the extracted power or cooling power requires finite-time thermodynamics, which relies on a microscopic model to describe the evolution of the system. Therefore, the maximization of the power is usually regarded as a model-specific task, thus lacking a universal characterization \cite{Alicki1979,Esposito2010,Abah2012, Zhang2014}. 

Conversely, the last decade has witnessed tremendous advances in experimental techniques \cite{Rossnagel2016,Josefsson2018,Ronzani2018,Maillet2019,Prete2019} which allow us to control quantum system and to operate them as thermodynamic machines \cite{Chen1994,Feldmann1996,Feldmann2000,Rezek2006,Arrachea2007,Scully2011,Abah2012, Correa2013,Dorfman2013,Brunner2014,Kosloff2014,Zhang2014,Campisi2015,Campisi2016,Cerino2016, Benenti2017,Brandner2017,Erdman2017, Suri2017,Watanabe2017,Cavina2018a,Erdman2018,Menczel2019a,Pekola2019,Bhandari2020}. We are now at the point that it is possible to fabricate devices which behave as qubits or qutrits, and couple them to thermal baths \cite{Pekola2015,Ronzani2018,Senior2020,VanHorne2020,Klatzow2019,VonLindenfels2019}. Typical experimental platforms, which range from trapped ions \cite{Friedenauer2008, Blatt2012}, to electron spins associated with nitrogen-vacancy centers \cite{Childress2006}, to circuit quantum electrodynamics \cite{Wallraff2004}, to single-electron transistors \cite{Kastner1992}, are all characterized by a set of ``control parameters'', e.g. electric or magnetic fields, that can be controlled in time by the experimentalist. The available control parameters may be subject to constraints, and may only grant us a partial control over the system dynamics. Given this framework, a fundamental question, which has not been tackled in general, is how to optimally drive the control parameters as to maximize the power of periodically driven classical or quantum thermal machines. This is the aim of the current paper.

In general, this is a formidable task, as it requires us to solve the time-dependent dynamics of an open quantum system, coupled to thermal baths, and to perform a functional optimization over all available control parameters. Within the slow-driving regime \cite{Esposito2010bis,Wang2011,Avron2012,Ludovico2016,Cavina2017,Abiuso2019}, a universal strategy to maximize the power has been recently derived \cite{Abiuso2020,abiuso2020geo}. Beyond this regime, common strategies to improve the power extracted from a quantum engine rely on performing fast and effectively adiabatic quantum operations through the Shortcut to Adiabaticity technique 
\cite{Deng2013,Torrontegui2013,Campo2014,Cakmak2018} or using Floquet engineering  \cite{Claeys2019,Villazon2019}.
The variety of frameworks employed span from the optimization 
of finite time Carnot cycles  \cite{Cavina2017,Abiuso2020, Allahverdyan2013, Dann2020} to Otto cycles \cite{Rezek2006,Quan2007,Abah2012,Karimi2016,Kosloff2017,Watanabe2017, Chen2019, Das2020} 
to endoreversible models \cite{Andresen1982,Song2006}.

However within this {\it mare magnum} of frameworks and methods, in the context of systems described by Markovian dynamics, recent evidence suggests that the optimal strategy to extract maximum power may consist of varying the control parameters \textit{infinitely fast} \cite{Geva1992, Cavina2018a, Erdman2019, Schmiedl2007,Pekola2019,cangemi2020optimal}. This observation would imply a profound ``duality'' between efficiency and power: both would be maximized according to two opposite universal strategies (infinitely slow, or infinitely fast control speed).

In the present paper we discuss the optimization of thermal machines in the fast driving regime.
This last, characterized by driving time scales which are much faster than the thermal relaxation, is introduced and discussed in the context of Markovian dynamics for systems whose Hamiltonian commutes at different times. This encompasses a variety of models of interest in stochastic thermodynamics, from chemical networks to molecular motors and more in general any dynamical system described by stochastic master equations \cite{seifert2012}.
Among all possible control strategies and protocols, we provide a universal proof that the power is optimized by ``generalized Otto cycles'', i.e. by performing sudden variations of the control parameters among a finite number of fixed values. We denote these sudden variations as ``quenches''. 
The generality of the proof is guaranteed by the fact that it holds for \textit{any} Hamiltonian describing the working fluid, the baths, and the coupling. Furthermore, it holds regardless of the number of baths, and regardless of the specific form of the time dependent dissipators in the Lindblad master equation, that can depend on an arbitrary number of external controls subject to arbitrary constraints.
In addition, it holds for the maximization of any linear combination of the heat currents, which includes the extracted power of a heat engine, the cooling power of a refrigerator, the dissipated heat by a heater, and so on.

The optimal protocol, i.e. the generalized Otto cycle, is characterized by $L$ infinitesimal time intervals, connected by an identical number of  quenches, in which the control parameters are held constant. We prove that, in general, $L \leq d$, where $d$ is the dimension of the Hilbert space of the working fluid. This bound also places a constraint on the number of thermal baths that are necessary to maximize the power. 

When all observables of the working fluid share the same (control-dependent) thermalization time, we further prove that $L=2$, that is, the optimal protocol is a standard infinitesimal Otto cycle.
In such models, assuming to have total control over the Hamiltonian of the working fluid, we identify the optimal modulation of the control parameters, which consists of producing a highly-degenerate many-body spectrum characterized by a single energy gap. 
This protocol allows us to compute the maximum achievable power using a working medium made up of $n$ interacting qubits.
We show that the power of such heat engine goes beyond its counterpart based on $n$ non-interacting qubits, displaying a many-body advantage. The value of the maximum power has a supra-extensive transient regime in $n$, and in the $n\rightarrow\infty$ limit we find that it is linear in the temperature difference $\Delta T$ between the hottest and the coldest bath, while the non interacting case exhibits the more common quadratic $\Delta T^2$ dependence. In addition, the interacting case displays an efficiency at maximum power which asymptotically approaches Carnot efficiency (for $n\to\infty$). Surprisingly, we find that in the refrigerator case, many-body interactions do not provide significant advantage over non-interacting qubits.

Next we study the qutrit system as a testing ground for our general results.
We numerically show that while the common $L=2$ case is optimal for typical thermalization models used to describe Bosonic and Fermionic baths, the generalized Otto cycle (characterized in this case by $L=3$ quenches) outperforms the $L=2$ case for some particular forms of the master equation. This implies that our bound on $L$ is, in general, tight. Furthermore, as opposed to the maximum efficiency, we show that the power can be enhanced by the presence of more than two thermal baths at different temperatures.

Our general result provides a solid characterization of optimal cycles for Markovian engines whose Hamiltonian commutes at different times. We conclude by discussing the non-commuting case, arguing that quantum non-adiabatic effects may produce different optimal cycles. This would represent an intriguing difference between semi-classical and quantum systems which deserves further investigation.

From an operational point of view, the results derived in this paper hugely simplify the numerical procedure of finding optimal protocols. Indeed, instead of having to optimize over all possible protocols, which are piece-wise continuous functions, using e.g. complex variational techniques \cite{Cavina2018b}, our results allow us to find the maximum power by optimizing a function of a fixed number of variables which is at most polynomial in the dimensionality of the Hilbert space of the working fluid. This is somewhat  analogous to what happens to control optimizations in the {\it slow driving} regime, in which the driving is much slower than the dissipative dynamics induced by the baths~\cite{Salamon1983,Cavina2017,Scandi2019,Abiuso2020,abiuso2020geo}. 
These results show that, by exploiting the concept of time scale separation, we can simplify the characterization of the power generation in thermal machines. 
%Dealing with the dynamics of ${\rho}(t)$ in intermediate regimes is in general more difficult, and this can be tackled, for instance, using the Pontryagin minimum principle technique~\cite{Kirk2004,Cavina2018a,Menczel2019a}.

The main results are ordered as follows.
In Sec.~\ref{sec:Poatm} we describe the theoretical model of a thermal machine used throughout the text, consisting of a quantum system coupled to an arbitrary number of Markovian thermal baths.
In Sec.~\ref{sec:fast_driving} we introduce and characterize the fast driving regime for periodically driven systems. We then prove the optimality of the generalized Otto cycle, and we discuss the bounds on the number of quenches $L$.
In Sec. \ref{sec:abiuso}  we apply the theory to a simple class of master equations, finding the exact form of the optimal driving protocols and highlighting the many-body advantage arising in this scenario. 
In Sec. \ref{sec:qutrit} we apply our general results to a qutrit thermal machine, in Sec.~\ref{sec:non-commute} we discuss the non-commuting case,
and in Sec.~\ref{sec:conclusions} we draw the conclusions.

 \begin{figure}[!htb]
	\centering
	\includegraphics[width=1\columnwidth]{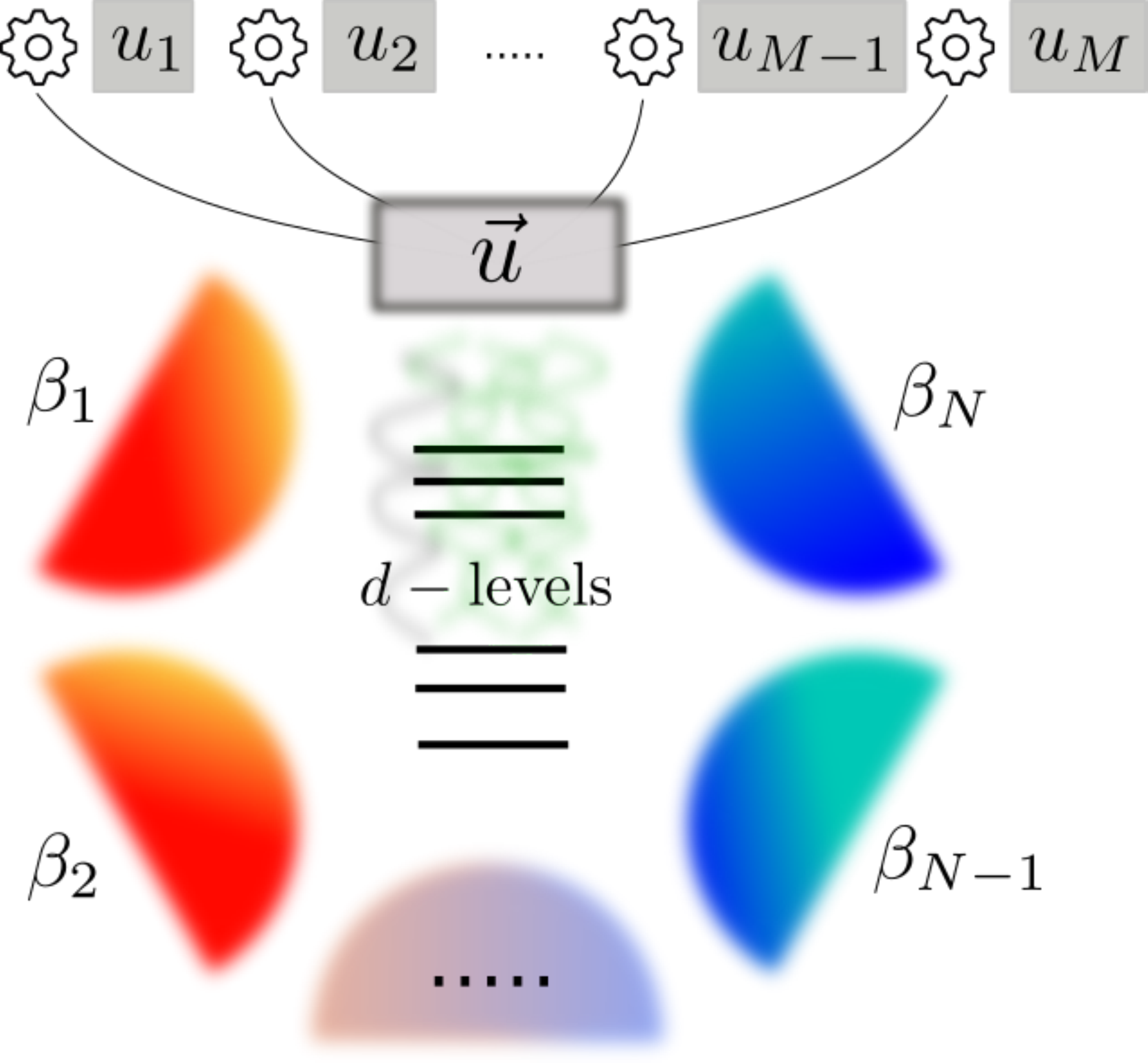}
	\caption{An arbitrary d-level system, controlled by $M$ parameters represented by the components of the vector $\vec{u}(t)$, is coupled to $N$ thermals baths.}
	\label{fig:main}
\end{figure}

\section{The model} 
\label{sec:Poatm}

As schematically depicted in Fig.~\ref{fig:main}, we consider a $d$-dimensional quantum system S 
(the working medium or working fluid of the model) that is 
 weakly coupled to  $N$ thermal baths characterized by inverse temperatures $\beta_\alpha$, for $\alpha=1,\dots ,N$.
We assume S to be externally controlled through a  set  of $M$ time-dependent control parameters  collectively  represented by a real vector function 
\begin{eqnarray} \label{control} 
\vec{u}(t): [0,\tau] \rightarrow \mathbb{D}\subseteq \mathbb{R}^M,\end{eqnarray}  where 
$\tau$ is the total duration of the driving, and $\mathbb{D}$ represents the set of the allowed
 values  the controls can assume, accounting for possible experimental constraints. In the following, we denote the function $\vec{u}(t)$ as the protocol or the driving. In our analysis $\vec{u}(t)$
acts as a modulator both for the  
  local Hamiltonian  of the system   ${{H}}_{\vec{u}(t)}$ as well as  for the interactions  with  the thermal baths which, 
 adopting the Gorini-Kossakowski-Sudarshan-Lindblad (GKSL) formalism~\cite{Gorini1976,Lindblad1976},
 we describe in terms of 
   the super-operator dissipators
 $\mathcal{D}_{\alpha, \vec{u}(t)}$.
We hence assign the temporal evolution of the system in terms of the following Master Equation (ME) for the
 reduced density matrix ${\rho}(t)$ of S, 
\begin{equation}
    {\partial_t}{{\rho}}(t) = \mathcal{L}_{\vec{u}(t)}\left[{\rho}(t)\right] \equiv
     -\frac{i}{\hbar}\left[{{H}}_{\vec{u}(t)}, {\rho}(t)\right] + \sum_{\alpha=1}^N \mathcal{D}_{\alpha, \vec{u}(t)}\left[{\rho}(t) \right]
    \label{eq:master_eq_rho},
\end{equation}
where $\mathcal{L}_{\vec{u}(t)}$ is the (time-dependent) quantum Liouvillian generator of the dynamics. 
Assuming that the Hamiltonian commutes at all times, i.e. that $[H_{\vec{u}_1},H_{\vec{u}_2}]=0$ for all $\vec{u}_1,\vec{u}_2 \in \mathbb{D}$, non-adiabatic transitions are not allowed, and the dissipators describe transitions between the instantaneous eigenstates of $H_{\vec{u}(t)}$. Therefore, $\mathcal{D}_{\alpha, \vec{u}(t)}$ only depends on time only through $\vec{u}(t)$, and not through the speed at which $\vec{u}(t)$ is modulated. In such regime, Eq.~(\ref{eq:master_eq_rho}) was shown to rigorously hold also in the driven case \cite{Davies1978,Grifoni1998,Yamaguchi2017,Dann2018}. We describe the possibility of deciding which bath is coupled to S at any given time through the dependence of the dissipators on $\vec{u}(t)$. If only bath $\alpha$ is coupled to S, and if we fix the control parameters $\vec{u}(t) = \vec{u}$, we expect S to thermalize by evolving towards the Gibbs density operator 
\begin{align}
\label{eq:gibbs_def}
    \rho^{\text{(eq)}}_{\alpha;\vec{u}} \equiv \exp[ -{\beta_\alpha} H_{\vec{u}}]/Z_{\alpha;\vec{u}}\ ,  
\end{align} 
$Z_{\alpha;\vec{u}} \equiv \mbox{Tr}[\exp[ -{\beta_\alpha} H_{\vec{u}}]]$ being the partition function. We frame this physical statement in mathematical terms by requiring all the dissipators $\mathcal{D}_{\alpha, \vec{u}}$
to be irreducible and adjoint-stable~\cite{Spohn1980,Menczel2019}, two conditions which, as we discuss in Appendix~\ref{app:projection}, are typically satisfied by non-pathological dissipators. 
The instantaneous heat flux flowing out of bath $\alpha$ can then be computed as \cite{Alicki1979} 
\begin{equation}
    {J}_\alpha(t) \equiv \Tr\left[  {{H}}_{\vec{u}(t)}\mathcal{D}_{\alpha, \vec{u}(t)}\left[{\rho}(t) \right] \right].
    \label{eq:heat_rho} 
\end{equation}

Within the above framework, we are interested in performing thermodynamic cycles, i.e. in performing a periodic driving $\vec{u}(t)$, with period $T$, such that the variation of internal energy 
\begin{equation}
	U(t) \equiv \Tr[H_{\vec{u}(t)}\rho(t)]
	\label{eq:u}
\end{equation}
of the working fluid is zero after each cycle. In this regime, the first law of thermodynamics guarantees us that all the work extracted from the system is only provided by the heat baths, and not by some particular state preparation of S. As we see from Eq.~(\ref{eq:u}), the periodicity of $U(t)$ requires both $\vec{u}(t)$ and $\rho(t)$ to be periodic functions. In general, $\rho(t)$ is not a periodic function. However, using the fact that the dissipators ${\cal D}_{\alpha,\vec{u}(t)}$ are irreducible and adjoint-stable, the Lindblad master equation enjoys the following property (a proof is provided by Theorem 2 of  Ref.~\cite{Menczel2019}): 
if  $\vec{u}(t)$ is a $T$-periodic function, then the solution of Eq.~(\ref{eq:master_eq_rho}) asymptotically 
converges toward a
``limiting cycle'' solution  $\rho_{[\vec{u}]}^{\text{(lc)}}(t)$, which is independent of the initial condition of the system, and which is periodic with the same period $T$ of the controls, i.e.   $\rho_{[\vec{u}]}^{\text{(lc)}}(t+T)=   \rho_{[\vec{u}]}^{\text{(lc)}}(t)$ for all~$t$
(the name ``limiting cycle'' follows from the fact that S naturally approaches it 
when   we repeat the periodic protocol ``many times'' \cite{Teschl2012}). The subscript in ${\rho}_{[\vec{u}]}^{\text{(lc)}}(t)$ emphasizes that the limiting cycle is a functional of the whole protocol, i.e. it depends on the control parameters along the whole cycle. In this asymptotic regime, the internal energy $U(t)$ becomes a periodic function, providing us with a thermodynamic cycle. From now on, we therefore focus solely on this regime.

We now wish to identify the optimal choice of $\vec{u}(t)$ that allows us to maximize the extracted power from a heat engine, or the cooling power of a refrigerator, averaged over a cycle. Both these quantities can be expressed as linear combinations of time integrals of the currents, defined in Eq.~(\ref{eq:heat_rho}).
Therefore, given an arbitrary collection $c_{\alpha}$ of real coefficients, we define the Generalized Average Power (GAP), which is a functional of the whole protocol, as 
\begin{multline}
    P_{\bold{c}}[\vec{u}] 
      \equiv \frac{1}{T} \sum_{\alpha=1}^N \int_0^{T}  c_{\alpha}{J}_{\alpha}(t) \, dt, \\ =
    \frac{1}{T} \int_0^{\tau} \Tr\left[  {{H}}_{\vec{u}(t)} \sum_\alpha c_\alpha \mathcal{D}_{\alpha, \vec{u}(t)}\left[{\rho}_{[\vec{u}]}^{\text{(lc)}}(t) \right] \right]dt.
    \label{eq:pa1} 
\end{multline}
 For instance, if we choose $c_{\alpha} = 1$ for all $\alpha$, Eq.~(\ref{eq:pa1}) represents the average of the total extracted heat flux, which coincides with the {\it average extracted power} for periodically driven heat engines; if instead $c_{\alpha} = \delta_{\alpha, N}$, with $\alpha=N$ labelling the coldest bath and $\delta$ representing the Kronecker delta,  Eq.~(\ref{eq:pa1}) represents the average {\it cooling power}, which measures the performance of a refrigerator; if $c_{\alpha} = -1$ for all $\alpha$, Eq.~(\ref{eq:pa1}) represents the average \textit{dissipated heat flux}, which measures the performance of a {\it heater}, and so on.

\section{Fast driving regime}
\label{sec:fast_driving}
Finding the optimal value of $\vec{u}(t)$ that 
maximize the functional~(\ref{eq:pa1}) 
is, in general, a formidable task. Nonetheless, as we shall see,  an explicit
solution to the problem can be obtained when studying the performance in the \textit{fast driving regime}. This is characterized by driving the system with a protocol $\vec{u}(t)$ whose period $T$ is much shorter than  the typical relaxation times induced by the baths. Therefore, we may expect that the limiting cycle state of S ``does not have time'' to thermalize with the bath, so it might actually converge to a fixed, time-independent out-of-equilibrium state. This is precisely what happens.

More specifically, let us denote with $\eta_{[\vec{u}]}$ the maximum rate which characterizes the ME~(\ref{eq:dyn2}) along the cycle, that is the rate characterizing the fastest possible relaxation to the steady state (see App.~\ref{bla} for a mathematical definition of $\eta_{[\vec{u}]}$).
Formally, we can expand ${\rho}^{\text{(lc)}}_{[\vec{u}]}(t)$ in a power series in $\eta_{[\vec{u}]} T\ll1$. As we prove in App.~\ref{bla}, it turns out that the leading order term ${\rho}_{[\vec{u}]}^{(0)}$ is indeed time-independent. A closed expression for such term can be obtained by making use of a projection technique that allows us to replace the dynamical generator $\mathcal{L}_{\vec{u}(t)}$ with the superoperator ${\cal G}_{\vec{u}(t)}$  which has the important property of being invertible on the $(d^2-1)$-dimensional linear subspace of traceless linear operators $\mathfrak{L}^{0}_{\text{S}}$ acting on S (see App.~\ref{app:projection} for details). Specifically, 
 Eq.~(\ref{eq:master_eq_rho}) can be rewritten in the more convenient form 
  \begin{equation}
 {\partial_t}{\tilde{\rho}}(t) =  {\cal G}_{\vec{u}(t)} \left[\tilde{\rho}^\text{(eq)}_{\vec{u}(t)}- \tilde{\rho}(t) \right],
 \label{eq:dyn2}  
\end{equation}
where ${\rho}^\text{(eq)}_{\vec{u}(t)}$ is the (unique) fixed point of  $\mathcal{L}_{\vec{u}(t)}$ and where, for all density matrices $\rho$ of S, we define 
\begin{eqnarray} \label{deftilde} \tilde{\rho} \equiv \rho - \openone/d\;, \end{eqnarray}  its traceless component. Equipped with this notation, in App.~\ref{bla} we prove that
 \begin{equation}
 \tilde{\rho}_{[\vec{u}]}^{(0)}
    \equiv \left(\int_{I_{[\vec{u}]}}  {\cal G}_{\vec{u}(t)} dt \right)^{-1}\left[\int_{I_{[\vec{u}]}}  {\cal G}_{\vec{u}(t)} [\tilde{\rho}^\text{(eq)}_{\vec{u}(t)}]\,dt  \right],
    \label{eq:p0}
\end{equation}
where $I_{[\vec{u}]}$ denotes the time interval of one cycle of duration $T$. The invertibility of $\int_{I_{[\vec{u}]}}  {\cal G}_{\vec{u}(t)} dt$ is guaranteed by the assumption that the dissipators are irreducible and adjoint-stable (see App.~\ref{app:projection} for details).
Using the approximation $\rho^\text{(lc)}_{[\vec{u}]}(t)\approx  {\rho}_{[\vec{u}]}^{(0)}$, we can write the GAP in Eq.~(\ref{eq:pa1}) in the fast driving regime as
\begin{equation}
    P_{\bold{c}}[\vec{u}] =
    \frac{1}{T} \int_{I_{[\vec{u}]}} \Tr\left[  {{H}}_{\vec{u}(t)} \sum_\alpha c_\alpha \mathcal{D}_{\alpha, \vec{u}(t)}\left[{\rho}_{[\vec{u}]}^{\text{(0)}} \right] \right]dt,
    \label{eq:pa_it} 
\end{equation}
which is guaranteed to be valid up to linear corrections in the expansion parameter $\eta_{[\vec{u}]} T$ 
(however, it should be stressed that, by direct evaluation, the GAP of the optimal protocol turns out to be valid up to second order corrections in $\eta_{[\vec{u}]} T$ in two level systems \cite{Cavina2018a,Erdman2019} and in the qutrit case studied in Sec.~\ref{sec:qutrit}). 

Equations (\ref{eq:p0}) and (\ref{eq:pa_it}) are the main starting point of our analysis: they allow us to express the GAP as an explicit functional of the protocol $\vec{u}(t)$ without requiring us to solve the ME.

\subsection{Optimality of sudden quenches}
\label{Oosq}
Instead of performing a direct constrained functional optimization of the GAP [see Eq.~(\ref{eq:pa_it})] with respect to $\vec{u}(t)$,   we will employ an iterative procedure that eventually leads to the identification of the ``generalized Otto cycle'' as the optimal one. The main idea of the proof is the following: given any assigned periodic protocol which respects the constraint $\vec{u}(t): [0,T] \rightarrow \mathbb{D}$, we prove that it is possible to ``cut away'' parts of it 
to build a new, shorter, cycle which delivers a higher or equal GAP than the starting one. By reiterating this process over and over, we end up with the generalized Otto cycle. We therefore denote this procedure as \textit{cut-and-choose}.

\begin{figure}[!tb]
	\centering
	\includegraphics[width=1\columnwidth]{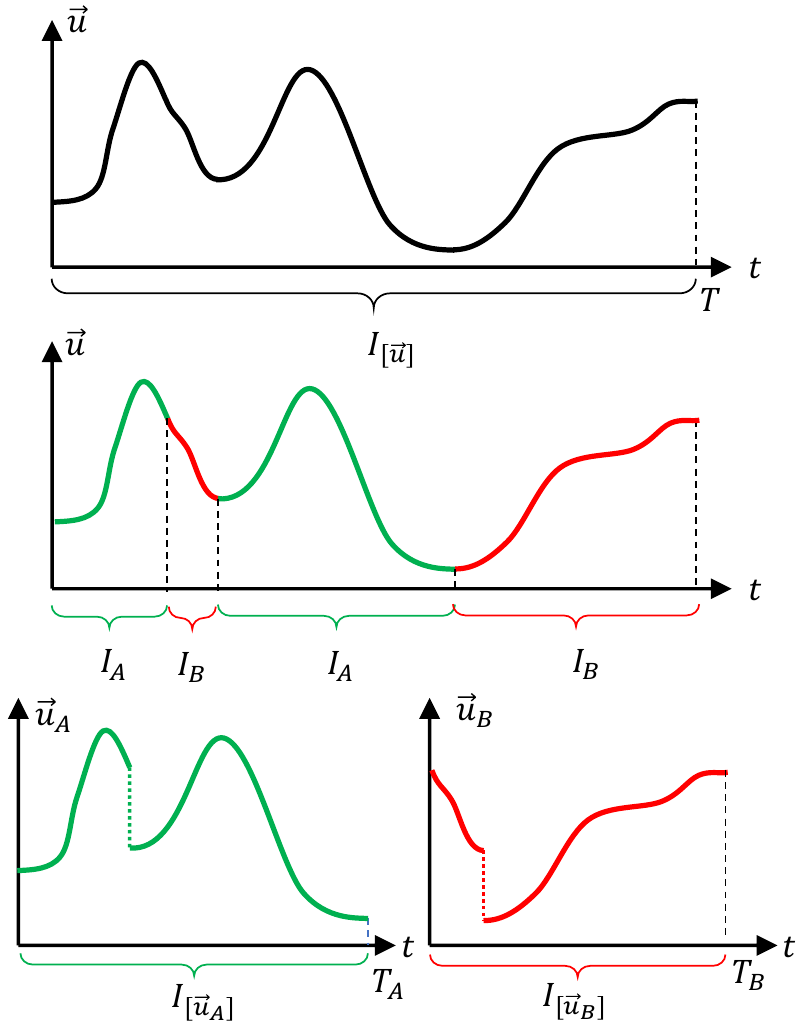}
	\caption{Schematic representation of the \textit{cut-and-choose} procedure for $M=1$. Upper panel: representation of an arbitrary protocol $\vec{u}(t)$ defined on the time interval $I_{[\vec{u}]}$ of duration $T$. Central panel: we partition $I_{[\vec{u}]}$ into two disjoing subsets $I_A$ and $I_B$. Lowe panel: we define two new sub-protocols $\vec{u}_A(t)$ and $\vec{u}_B(t)$ by restricting $\vec{u}(t)$ respectively to $I_A$ and $I_B$. This process may introduce discontinuities in the controls, denoted as quenches. }
	\label{fig:dei_1d}
\end{figure}

In order to detail the \textit{cut-and-choose} procedure, let us first formally introduce the notion of cyclic sub-protocols.
Given an arbitrary cyclic protocol $\vec{u}(t)$ of period $T$ and fundamental period
 $I_{[\vec{u}]} = [0,T]$,  consider a subset $I_A$ of  $I_{[\vec{u}]}$ of non-zero measure $T_A$. 
  A cyclic sub-protocol $\vec{u}_A(t)$ of $\vec{u}(t)$  with period $T_A$ and fundamental period $I_{[\vec{u}_A]}\equiv [0, T_A]$ is hence obtained  by rigidly joining  the various parts which compose the 
  restriction of $\vec{u}(t)$ on $I_A$. This procedure may introduce localized discontinuities, i.e. quenches, within the protocol -- see Fig.~\ref{fig:dei_1d} for an example for $M=1$.
Assume now to drive S by repeating many times the selected sub-protocol:
since the image points of the curve $\vec{u}_A(t):
I_{[\vec{u}_A]} \rightarrow  \mathbb{D}$ form a proper subset of those of $\vec{u}(t): I_{[\vec{u}]} \rightarrow  \mathbb{D}$, it follows that 
 if   the fast driving limit holds for the latter, i.e. if $\eta_{[\vec{u}]} T\ll1$, then the same condition applies also to $\vec{u}_A(t)$, i.e. $\eta_{[\vec{u}_A]} T_A\ll1$ -- see Appendix~\ref{newappa}. Furthermore, by construction, the new cyclic sub-protocol satisfies the constraints on the values of the control.

Since Eq.~(\ref{eq:p0}) holds for any periodic protocol in the fast driving regime, by repeating $\vec{u}_A(t)$ many times, the state of S will tend to a new asymptotic constant state  ${\rho}^{(0)}_{[\vec{u}_A]}$ whose
  traceless component reads
    \begin{equation}
 \tilde{\rho}_{[\vec{u}_A]}^{(0)}
    = \left(\int_{I_{[\vec{u}_A]}}   {\cal G}_{\vec{u}_A(t)} dt \right)^{-1}\left[\int_{I_{[\vec{u}_A]}}  {\cal G}_{\vec{u}_A(t)} [\tilde{\rho}^\text{(eq)}_{\vec{u}_A(t)}]\,dt \right].
    \label{eq:p0A}
\end{equation}
It goes without mentioning that analogous conclusions can be drawn also for the
sub-protocol $\vec{u}_B(t)$ that is obtained by considering the restriction of $\vec{u}(t)$ to the complement $I_B$ of $I_A$, i.e. the set $I_B = I_{[\vec{u}]}/ I_A$ of measure $T_B = T-T_A$: 
 once more, under iterated application of such driving, the state of S will tend to a constant asymptotic state ${\rho}^{(0)}_{[\vec{u}_B]}$ given by Eq.~(\ref{eq:p0A}) by simply replacing everywhere the index $A$ with~$B$; see Fig.~\ref{fig:dei_1d} for an example.

\begin{figure}[!tb]
	\centering
	\includegraphics[width=1\columnwidth]{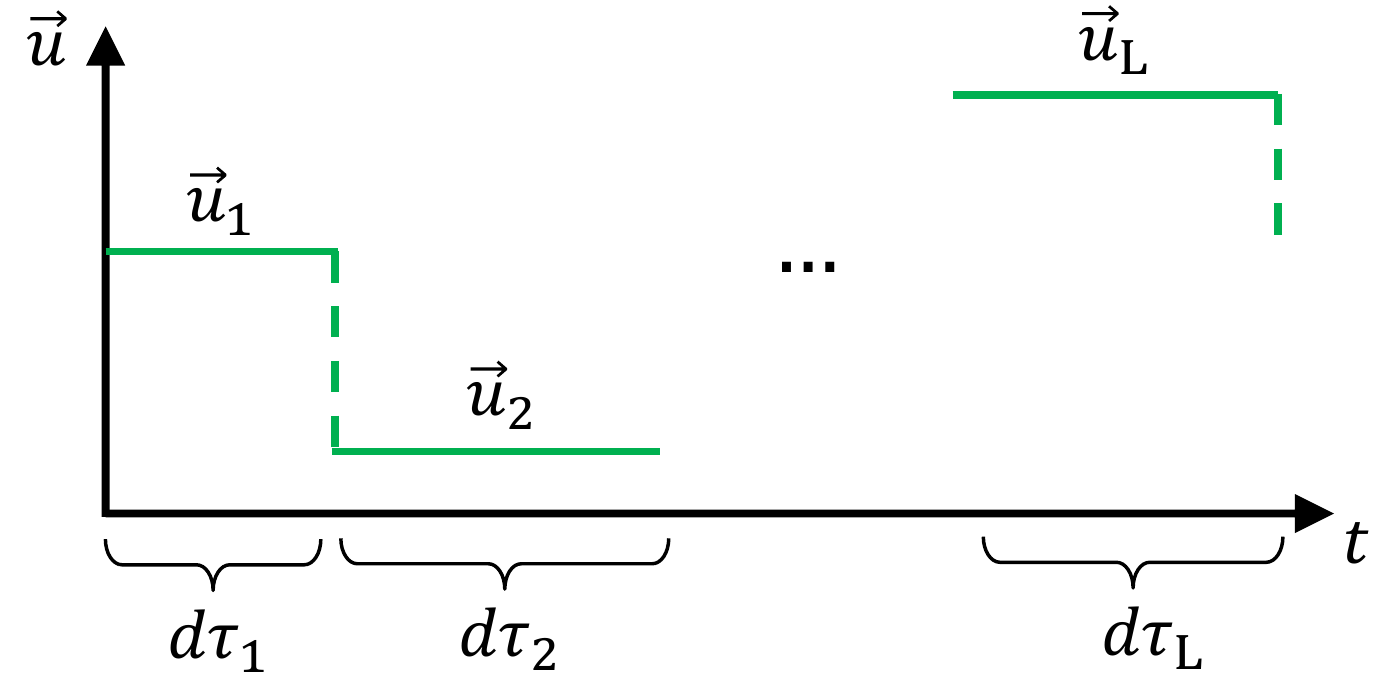}
	\caption{Representation of a generalized Otto cycle, which results from a large number of re-iterations of the \textit{cut-and-choose} procedure depicted in Fig.~\ref{fig:dei_1d}. $L$ (infinitesimally short) constant controls are alternated with quenches. Finite upper bounds can be obtained for the maximum value of $L$ that is needed for the optimization (see Table~\ref{table:params}).}
	\label{fig:gen_otto}
\end{figure}

Assume next that the states ${\rho}^{(0)}_{[\vec{u}_A]}$ and ${\rho}^{(0)}_{[\vec{u}_B]}$
introduced above coincide and are equal to ${\rho}^{(0)}_{[\vec{u}]}$, i.e.
\begin{eqnarray} \label{STRONG} 
{\rho}^{(0)}_{[\vec{u}_A]} = {\rho}^{(0)}_{[\vec{u}_B]}={\rho}^{(0)}_{[\vec{u}]}\;.
\end{eqnarray} 
Equation~(\ref{STRONG})  is a rather strong requirement which in general is not
met by generic choices of $I_A$ and $I_B$: still, as we shall discuss in the next section, the possibility of identifying sub-protocols fulfilling this property is always granted. For the moment we hence assume that Eq.~(\ref{STRONG}) is satisfied. The GAPs $P_{\bold{c}}[\vec{u}_A]$ and $P_{\bold{c}}[\vec{u}_B]$ delivered respectively by the sub-protocols $\vec{u}_A$ and $\vec{u}_B$ can be computed using Eq.~(\ref{eq:pa_it}). Assuming Eq.~(\ref{STRONG}) is fulfilled, we notice that the integrands entering $P_{\bold{c}}[\vec{u}]$, $P_{\bold{c}}[\vec{u}_A]$ and $P_{\bold{c}}[\vec{u}_B]$ are all the same. Therefore, exploiting the linearity of the integral respect to the its integration domain (i.e. time), and recalling that $T= T_A + T_B$, we have that
 \begin{eqnarray} \label{CONVEX} 
 P_{{\bf c}}[\vec{u}] = \frac{T_A P_{{\bf c}}[\vec{u}_A] + T_B P_{{\bf c}}[\vec{u}_B]}{T_A + T_B} .
 \end{eqnarray}  
  The above equation establishes that  the GAP of the original protocol $\vec{u}(t)$ can be expressed as a  non-trivial convex combination of the GAPs of the sub-protocols  $\vec{u}_A(t)$ and $\vec{u}_B(t)$: therefore it must be smaller or equal to the maximum of those two quantities, i.e. 
\begin{equation}
    P_{{\bf c}}[\vec{u}] \leq P_{{\bf c}}[\vec{u}_A],
    \label{eq:pa_magg}
\end{equation}
where, without loss of generality we assumed 
$P_{{\bf c}}[\vec{u}_B]\leq P_{{\bf c}}[\vec{u}_A]$.
Inequality (\ref{eq:pa_magg}) implies that given a generic periodic protocol $\vec{u}(t)$, it is 
possible to construct a \textit{shorter} one $\vec{u}_A$ that delivers a larger or equal GAP. This is the reason for the name \textit{cut-and-choose} procedure.
We can now re-iterate the \textit{cut-and-choose} procedure starting from $\vec{u}_A(t)$, thus obtaining another (even shorter) protocol $\vec{u}_{AA}(t)$ that produces a greater or equal GAP, and so on and so forth.
After many iterations of the \textit{cut-and-choose} procedure, we end up with a protocol that cannot be further optimized via this technique. This protocol is characterized by an infinitesimal domain $I$ of duration $d\tau$, divided into $L$ segments of length $d\tau_i$. Without loss of generality, we can assume that the $d\tau_i$s are short enough such that the controls $\vec{u}(t)$ take on a constant value $\vec{u}_i \equiv \vec{u}(t_i) \in \mathbb{D}$ during each time interval $d\tau_i$. This is a generalized Otto cycle; see Fig.~\ref{fig:gen_otto} for a schematic representation.

Using Eq.~(\ref{eq:pa_it}), the associated GAP of such protocol can hence be expressed as
\begin{equation} 
 P_{{\bf c}}[\{\vec{u}_i,\mu_i\}]  =  \sum_{j=1}^L \mu_j \Tr\left[  {{H}}_{\vec{u}_j} \sum_\alpha c_\alpha \mathcal{D}_{\alpha, \vec{u}_j}\left[{\rho}_{[\{\vec{u}_i,\mu_i\}]}^{\text{(0)}} \right] \right],
 \label{eq:gap_otto}
\end{equation}
where $\mu_i= d\tau_i/d\tau$ represents the percentage of the total protocol time spent at each point $\vec{u}_i$, and ${\rho}^{(0)}_{[\{\vec{u}_i,\mu_i\}]}$ is the time-independent limiting cyclic state whose traceless component is [see Eq.~(\ref{eq:p0})]
\begin{equation}
   \tilde{\rho}^{(0)}_{[\{\vec{u}_i,\mu_i\}]}     \equiv \left(\sum_{j=1}^L \mu_j{\cal G}_{\vec{u}_j}   \right)^{-1} \left[ \sum_{j=1}^L \mu_j {\cal G}_{\vec{u}_j}[ \tilde{\rho}^\text{(eq)}_{\vec{u}_j}]\right]\;.
    \label{eq:p0_otto}
\end{equation}

Crucially, we are able to place a finite upper bound to the number $L$ of time intervals of the optimal generalized Otto cycle. In App.~\ref{app:max_quench} we prove that, to maximize the GAP in general, it is sufficient to consider $L$ to be at most equal to the degrees of freedom of the 
density matrix plus one. Since in the commuting case only the diagonal component of $\rho(t)$ plays a role in determining the heat currents (see App.~\ref{appfast} for details), we find that $L\leq d$. Furthermore, as discussed in Sec.~\ref{sec:abiuso}, if the dissipator of each bath is characterized by a single (control-dependent) timescale and $N=2$, then $L=2$ regardless of the dimensionality of the system; in this case, the optimal protocol reduces to a conventional infinitesimal Otto cycle.

We proved that the generalized Otto cycle universally maximizes the GAP. We can thus directly optimize Eq.~(\ref{eq:gap_otto}) over the values of the controls
$\vec{u}_i$ and of the time fractions $\mu_i$ which are model-specific.
The total number of scalar parameters over which Eq.~(\ref{eq:gap_otto}) must be optimized is given by $(L-1) + ML$, where $L-1$ comes from the choices of fractions  $\mu_i$, and $ML$ is the number of scalar control parameters. We report a summary of these results in Table~\ref{table:params}.

%%%%%%
\begin{table}[tb!]
\centering
\begin{tabular}{|c|c|c|}
	\hline
	~ &  General & Simple relax. \\
	\hline\hline
	$\max \, L$ &  $d$ & $2$\red{$^*$} \\
	\hline
	Scalar Parameters & $d(M+1) - 1$ &  $2$\red{$^*$}$(M + 1)-1$ \\
	\hline
\end{tabular}
\caption{Maximum value of the number of time fractions $L$ and of the scalar parameters $M(L+1)-1$ which determine the generalized Otto cycle
that optimise the engine performances in the general case and for a simple choice of the Lindbladian form (cfr. Sec.\ref{sec:abiuso}). Here $d$ is the dimension of the working fluid and $M$ the number of components of the external control $\vec{u}(t)$. \\
\red{*}: The value 2 holds for refrigerators with any number of thermal sources, or for a
%while this value is $\min(N,d)$ in the 
heat engines with 2 thermal reservoirs. See  Sec.~\ref{sec:abiuso} for details.}
\label{table:params}
\end{table}
%%%%%%

 In order to gain further physical insight into our result, let us consider the paradigmatic case in which our system S can only be coupled to one bath at the time. Mathematically, this assumption can be described by a specific control parameter, say $\alpha(t)$, whose value is the index of the bath we are coupled to, $\alpha=1,\dots, N$. Therefore, 
 S must be coupled only to a single bath in each time interval $d\tau_i$.
In this scenario, it is interesting to notice that our bound on the number of time intervals poses a limit to the maximum number of thermal baths necessary to maximize the GAP: indeed, at most $L$ baths will be used. Therefore, for low dimensional working fluids, the maximum number of thermal baths necessary to maximize the GAP is strongly limited. However, we also explicitly show in Sec.~\ref{sec:qutrit} that three thermal baths at different temperatures can outperform two thermal baths when the working fluid is a qutrit. This result is in contrast with the maximization of the efficiency, which is always obtained by coupling S only to the hottest and coldest bath available.

As a final technical remark, we discuss how to simplify the optimization over the choice of the bath coupled to S. In principle, any bath can be coupled to S during each time interval $d\tau_i$.
However, by direct inspection of Eqs.~(\ref{eq:gap_otto}) and (\ref{eq:p0_otto}), it can be seen that the GAP is invariant under permutations of $\vec{u}_i$ and $\mu_i$. Therefore, the number of independent choices of the bath is given by the binomial coefficient ${L+N-1\choose N-1}$ which e.g. scales linearly in $L$ when only two thermal baths are available. The maximization of the GAP is thus carried out by repeating the optimization of Eq.~(\ref{eq:gap_otto}) over the other control parameters for each independent choice of the baths, and then choosing the configuration delivering the largest GAP.

\subsection{A geometric interpretation of Eq.~(\ref{STRONG}) } 
\label{sec:average_partition}

The argument presented in the previous section  relies on the assumption~(\ref{STRONG})
that one can identify 
two new sub-protocols $\vec{u}_A(t)$ and $\vec{u}_B(t)$ that preserve the asymptotic state ${\rho}^{(0)}_{[\vec{u}]}$ of the original protocol $\vec{u}(t)$. 
We provide an explicit proof that   
such condition can always be fulfilled by 
translating it into a geometric problem.

\begin{figure}[!tb]
	\centering
	\includegraphics[width=1\columnwidth]{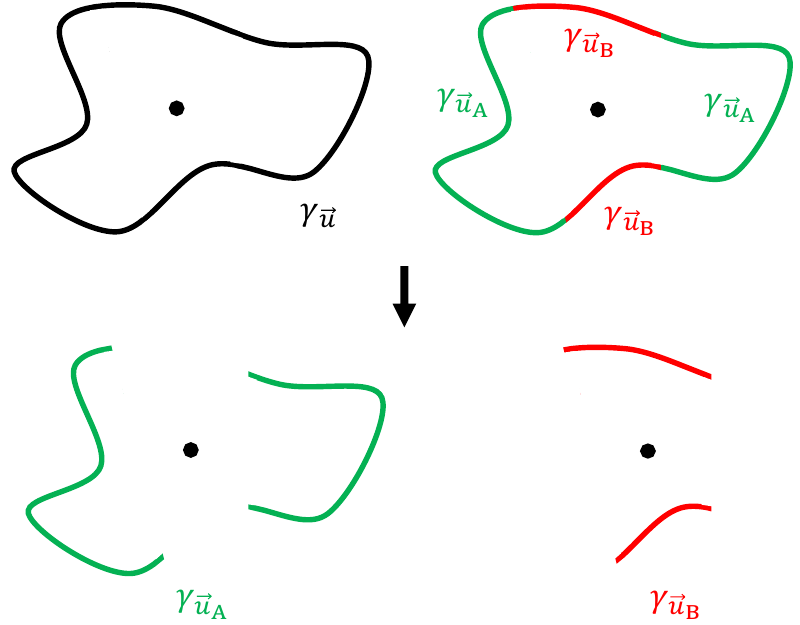}
	\caption{Upper left: schematic representation of $\gamma_{[\vec{u}]}$ and of its center of mass. Upper right: schematic representation of a partition of $\gamma_{[\vec{u}]}$ into $\gamma_{[\vec{u}]_A}$ and $\gamma_{[\vec{u}]_B}$. Lower panels: schematic representation of the sub-curves  $\gamma_{[\vec{u}]_A}$ and  $\gamma_{[\vec{u}]_B}$ which preserve the center of mass of the original curve $\gamma_{[\vec{u}]}$. }
	\label{fig:dei_2d}
\end{figure}

For this purpose, let us define the curves 
 $\gamma_{[\vec{u}]}\equiv \{  {v}_{\vec{u}(t)} | t \in I_{[\vec{u}]}\}$, 
$\gamma_{[\vec{u}_A]}\equiv \{  {v}_{\vec{u}_A(t)} | t \in I_{[\vec{u}_A]}\}$, and 
$\gamma_{[\vec{u}_B]}\equiv \{  {v}_{\vec{u}_B(t)} | t \in I_{[\vec{u}_B]}\}$
generated by the   functions
\begin{eqnarray}
{v}_{\vec{u}(t)}&\equiv& {\cal G}_{\vec{u}(t)} \left[ \tilde{\rho}^\text{(eq)}_{ \vec{u}(t)} - \tilde{\rho}^{(0)}_{[\vec{u}]} \right] \;, \\ 
   {v}_{\vec{u}_{A,B}(t)} &\equiv& {\cal G}_{\vec{u}_{A,B}(t)} \left[ \tilde{\rho}^\text{(eq)}_{ \vec{u}_{A,B}(t)} - \tilde{\rho}^{(0)}_{[\vec{u}]} \right]\,.
\label{vecdef}
\end{eqnarray} 
Since the domains $I_{[\vec{u}_{A}]}$ and $I_{[\vec{u}_{B}]}$ are complementary and provide a decomposition of $I_{[\vec{u}]}$, $\gamma_{[\vec{u}_A]}$ and $\gamma_{[\vec{u}_B]}$ are disjoint, and their union coincides with $\gamma_{[\vec{u}]}$ (see upper panels of Fig.~\ref{fig:dei_2d} for a schematic representation). 
It is important to notice that the functions in Eq.~(\ref{vecdef}), thus also the curves $\gamma_{[\vec{u}]}$, $\gamma_{[\vec{u}_A]}$, and $\gamma_{[\vec{u}_B]}$,
belong to the special subspace of  $\mathfrak{L}^{0}_{\text{S}}$  formed by the traceless Hermitian operators of S which is locally isomorphic to $\mathbb R^{D}$, with $D\equiv d^2-1$.% can be parameterized in terms of $d^2-1$ real parameters. 
 By exploiting the fact that the Hamiltonian commutes at all times, we can further reduce the number of degrees of freedom to $D\equiv d-1$. This is due to the fact that the heat currents can be written solely in terms of the diagonal part of $\rho(t)$, which in turn satisfies a closed equation of motion (see App.~\ref{appfast} for details). 
%We can thus consider the subspace of \textit{diagonal} traceless operators of S, which are isomorphic to $R^{d-1}$. 

Since $\tilde{\rho}^{(0)}_{[\vec{u}]}$ satisfies Eq.~(\ref{eq:p0}), the curve
 $\gamma_{[\vec{u}]}$  has a null  ``center of mas'' $O_{[\vec{u}]}$ (represented by the black dot in Fig.~\ref{fig:dei_2d}), i.e. 
 \begin{eqnarray} 
	O_{[\vec{u}]} \equiv   \int_{I_{[\vec{u}]}} {v}_{\vec{u}(t)}\, dt=0\;.
\end{eqnarray} 
Using the linearity of the integral respect to its integration domain, it is easy to verify that the sum of the 
 ``centers of mass'' $O_{[\vec{u}_{A}]} \equiv  \int_{I_{[\vec{u}_{A}]}} {v}_{\vec{u}_{A}(t)}\, dt$ with $O_{[\vec{u}_{B}]} \equiv  \int_{I_{[\vec{u}_{B}]}} {v}_{\vec{u}_{B}(t)}\, dt$ is null, i.e.
\begin{eqnarray} \label{idd} 
O_{[\vec{u}_{A}]} + O_{[\vec{u}_{B}]}= O_{[\vec{u}]}= 0\;.
\end{eqnarray} 
We claim that  a necessary and sufficient  condition for Eq.~(\ref{STRONG}) to hold
is that the curve $\gamma_{[\vec{u}_A]}$ (and hence 
due to Eq.~(\ref{idd}), also
 $\gamma_{[\vec{u}_B]}$) must have a null center of mass too. Indeed, exploiting the invertibility of $\int_{I_{[\vec{u}_A]}}  {\cal G}_{\vec{u}_A(t)} dt$ on $\mathfrak{L}^{0}_{\text{S}}$, one can
  observe that 
setting $O_{[\vec{u}_{A}]} =0$ is fully equivalent to having
\begin{multline}  
	\tilde{\rho}_{[\vec{u}]}^{(0)}
= \left(\int_{I_{[\vec{u}_A]}}  {\cal G}_{\vec{u}_A(t)} dt \right)^{-1}\left[\int_{I_{[\vec{u}_A]}}  {\cal G}_{\vec{u}_A(t)} [\tilde{\rho}^\text{(eq)}_{\vec{u}_A(t)}]\,dt 
   \right] 
  \\ = \tilde{\rho}_{[\vec{u}_A]}^{(0)},
\end{multline}
where, in the last step, we used Eq.~(\ref{eq:p0A}) to recognize $\tilde{\rho}_{[\vec{u}_A]}^{(0)}$. An analogous conclusion holds also for the sub-protocol $\vec{u}_B(t)$ thanks to Eq.~(\ref{idd}).

This is the geometric reformulation of Eq.~(\ref{STRONG}) we were looking for: our partitioning technique works if, starting from a generic curve $\gamma_{[\vec{u}
]}$ in $\mathbb R^D$ having a null center of mass, we are able to split it into two sub-curves $\gamma_{[\vec{u}_A]}$ and $\gamma_{[\vec{u}_B]}$ such that these still have a null center of mass  (this concept is schematically represented in Fig.~\ref{fig:dei_2d}).
In Appendix \ref{app:infinitesimal} we prove that it is indeed possible assuming that the original protocol  $\vec{u}(t)$ possesses some weak notion of regularity.
The main idea is that, given an arbitrary curve in $\mathbb{R}^{D}$ with zero center of mass, it is always possible to identify a null convex combination of at most $D+1$ points lying on the curve.
For sufficiently regular curves, the implicit function theorem allows us to extend these points to a piecewise continuous curve of finite size. 

Reiterating this \textit{cut-and-choose} procedure many times may lead to a piecewise continuous curves with a large number of discontinuities. Crucially, in App.~\ref{app:max_quench} we show that it is always possible to end up with a curve characterized by at most $D+1$ discontinuities. This result gives rise to the bounds on $L$ summarized in Table~\ref{table:params}.

\section{Simple relaxation case}
\label{sec:abiuso}
In this section we discuss a simplified model of thermalization
where  the super-operator ${\cal G}_{\vec{u}(t)}$ of Eq.~(\ref{eq:dyn2}) 
is purely multiplicative, leading to a ME of the form 
\begin{equation} 
\label{eq:scalar_dynamics}
{\partial_t}{\tilde{\rho}(t)} ={\Gamma}_{\vec{u}(t)}(\tilde{\rho}^\text{(eq)}_{\vec{u}(t)} - \tilde{\rho}(t)), 
\end{equation}
with ${\Gamma}_{\vec{u}(t)} > 0$ 
a scalar number which defines the rate of thermalization of all the observables of the system. 
Furthermore, we assume that the model allows S to be coupled to a single bath at the time. As discussed in the final part of Sec.~\ref{Oosq}, we formally introduce a single control parameter, denoted with $\alpha(t)$, indexing the bath we are coupled to at time $t$. Notice that, for all values of $\vec{u}(t)$, the equilibrium states ${\rho}^\text{(eq)}_{\vec{u}(t)}$ always correspond to the Gibbs distribution of bath $\alpha(t)$, i.e. $\rho^{\text{(eq)}}_{\alpha(t);\vec{u}(t)}$ as in Eq.~\eqref{eq:gibbs_def}. 
As discussed in the first part of the paper, the maximum GAP is given by Eqs.~(\ref{eq:gap_otto}) and (\ref{eq:p0_otto}) which, using Eq.~(\ref{eq:scalar_dynamics}), can be rewritten as 
\begin{eqnarray} 
     P_{\bf c}[\{\vec{u}_i,\mu_i\}]&=&\dfrac{\sum_{i,j=1}^L c_{\alpha_i} \mu_i \mu_j  \Gamma_{\vec{u}_i} \Gamma_{\vec{u}_j}
P_{i\leftarrow j}
}{\sum_{i=1}^L \mu_i  \Gamma_{\vec{u}_i}}\;, 
\label{eq:pow_simple} \\
\label{eq:rho0_simple}
 \tilde{\rho}^{(0)}_{[\{\vec{u}_i,\mu_i\}]}
    &=& \dfrac{\sum_{i=1}^L \mu_i  \Gamma_{\vec{u}_i}
     \tilde{\rho}^\text{(eq)}_{\vec{u}_i}}{\sum_{i=1}^L \mu_i  \Gamma_{\vec{u}_i} }\;, 
\end{eqnarray}
where $\alpha_i$ is the constant value of $\alpha(t)$ during the interval $d\tau_i$ (we used the assumption that S can be coupled to a single bath at the time to remove the sum over $\alpha$ in Eq.~(\ref{eq:pow_simple})),
  and where
\begin{equation}
\label{def:Pij}
P_{i\leftarrow j}\equiv\Tr\left[{{H}}_{\vec{u}_i}\left(\tilde{\rho}^{\rm (eq)}_{\vec{u}_i}-\tilde{\rho}^{\rm (eq)}_{\vec{u}_j}\right)\right]\;.
\end{equation}
Notice that $P_{i\leftarrow i}=0$, while 
\begin{eqnarray} 
\label{eq:Psym_property}
P_{j\leftarrow k}+P_{k\leftarrow j}\leq 0\;,
\end{eqnarray}           
when S is coupled to the same temperature during the time intervals $d\tau_j$ and $d\tau_k$. This is given by the fact that  $P_{j\leftarrow k}+P_{k\leftarrow j}$ is equal to $ P_{\bf c}[\{\vec{u}_i,\mu_i\}]$ with  $c_\alpha=1$ $\forall i$, and $\mu_i=0$ $\forall i\neq j,k$, which physically represents the average power extracted from a heat engine operating between equal temperatures (and therefore cannot be positive)~\cite{NOTA}.

Using these properties, we show that with the only assumption of the dynamics being described by~ Eq.~\eqref{eq:scalar_dynamics}, it is possible to greatly simplify the optimization of the GAP of thermal machines. 
As shown in Appendix~\ref{app:positive_GAPs}, we consider \textit{positive} GAPs, i.e. generalized average powers consisting of a positive linear combination of the heat currents extracted from the different thermal baths (formally we assume that  $c_\alpha \geq 0$ $\forall \alpha$). This hypothesis includes both the average power extracted from a heat engine ($c_\alpha=1$ $\forall \alpha$), and the cooling power of a refrigerator ($c_\alpha=\delta_{\alpha,N}$, with $\alpha=N$ labelling the coldest bath). We prove that, in order to maximize a positive GAP, it is sufficient to consider a protocol with at most one time interval per temperature; therefore $L\leq N$. Moreover, if more than one heat current is neglected in the definition of the GAP, it is possible to further reduce the number of intervals. 
Specifically, given $\kappa\leq N$ the number of distinct temperatures of the baths for which $c_\alpha\neq 0$, we prove that
\begin{equation}
L\leq \min(N,\kappa+1)\ .
\end{equation}
This implies that a refrigerator ($\kappa=1$) is always characterized by $L=2$, regardless of the number of baths, while a heat engine ($\kappa=N$) by $L\leq N$. In the following, for simplicity, we focus on the refrigerator and heat engine case with two thermal baths at our disposal. As a consequence, $L=2$. Under this hypothesis, we find that:
\begin{itemize}
\item[(i)] The optimal durations of the time intervals and the resulting maximum GAPs can be determined as a function of the control parameters (Sec.s~\ref{subsec:simple_refrig}-\ref{subsec:simple_engine}).
\item[(ii)] If the thermalization rates are a function of the bath $\alpha(t)$, but only weakly depend on the specific value of the other control parameters, i.e. $\Gamma_{\vec{u}_i} = \Gamma_{\alpha_i}$, and if we assume to have total control over the system Hamiltonian, we can fully carry out the maximization of the GAP, finding that the optimal control strategies involve degenerate spectra of the Hamiltonian of the working fluid (Sec.~\ref{subsec:spectra_opt}).
\item[(iii)] Under the toy model hypothesis of (ii), we compare the GAP of a heat engine and of a refrigerator delivered by $n$ non-interacting qubits [$\text{GAP}_\text{NI}(n)$], with the GAP of $n$ interacting qubits [$\text{GAP}_\text{I}(n)$], finding that there is a many-body advantage in the engine case.
\end{itemize}

\subsection{Refrigerator}
\label{subsec:simple_refrig}
Let us consider two inverse temperatures $\beta_1$ and $\beta_2$ such that $\beta_1 < \beta_2$. The average cooling power of a refrigerator, $P_\text{[R]}$, is described by the GAP with $c_2 = 1$ on the cold bath while $c_1=0$. Since $L=2$, Eq.~\eqref{eq:pow_simple} reduces to
\begin{eqnarray}
\label{eq:pow_simple_fridge}
P_\text{[R]}= \;  \frac{\mu_1  \mu_2  \Gamma_{\vec{u}_1}\Gamma_{\vec{u}_2} P_{2\leftarrow 1}}{\mu_1  \Gamma_{\vec{u}_1} +\mu_2  \Gamma_{\vec{u}_2}} \;
\end{eqnarray}
where $\mu_2= 1-\mu_1$. 
We can thus explicitly maximize the above expression over the choice of the time fraction $\mu_1$, leading to
\begin{equation}
\label{eq:pow_fridge_max}
P^{\rm (max)}_\text{[R]}=\dfrac{\Tr\left[{H}_{\vec{u}_2}\left(\tilde{\rho}^{\rm (eq)}_{\vec{u}_2}-\tilde{\rho}^{\rm (eq)}_{\vec{u}_1}\right)\right]}{\left(\sqrt{\Gamma_{\vec{u}_1}^{-1}}+\sqrt{\Gamma_{\vec{u}_2}^{-1}}\right)^2},
\end{equation}
which is obtained for $\mu_1=\sqrt{\Gamma_{\vec{u}_2}}/(\sqrt{\Gamma_{\vec{u}_2}}+\sqrt{\Gamma_{\vec{u}_1}})$. Notably, the expression of the maximum cooling power in Eq.~(\ref{eq:pow_fridge_max}) only requires a maximization over $\vec{u}_1$ and $\vec{u}_2$, which in general is model dependent.

\subsection{Engine}
\label{subsec:simple_engine}
Let us consider the same setting $\beta_1 < \beta_2$. The average extracted power of a heat engine, $P_\text{[E]}$, is described by the GAP with $c_1=c_2=1$. Since $L=2$, Eq.~\eqref{eq:pow_simple} reduces to
\begin{equation}
\label{eq:pow_simple_engine}
P_\text{[E]}= \;  \frac{\mu_1  \mu_2  \Gamma_{\vec{u}_1}\Gamma_{\vec{u}_2} \left(P_{1\leftarrow 2}+P_{2\leftarrow 1}\right)}{\mu_1  \Gamma_{\vec{u}_1} +\mu_2  \Gamma_{\vec{u}_2}} \;
\end{equation}
It follows that the optimization over the time fraction $\mu_1$ is identical to that of the refrigerator, see Eq.~\eqref{eq:pow_simple_fridge}, leading to
\begin{equation}
\label{eq:pow_engine_max}
P^{\rm (max)}_\text{[E]}=\dfrac{\Tr\left[\left({H}_{\vec{u}_1}-{H}_{\vec{u}_2}\right)\left(\tilde{\rho}^{\rm (eq)}_{\vec{u}_1}-\tilde{\rho}^{\rm (eq)}_{\vec{u}_2}\right)\right]}{\left(\sqrt{\Gamma_{\vec{u}_1}^{-1}}+\sqrt{\Gamma_{\vec{u}_2}^{-1}}\right)^2},
\end{equation}
which is obtained for $\mu_1=\sqrt{\Gamma_{\vec{u}_2}}/(\sqrt{\Gamma_{\vec{u}_2}}+\sqrt{\Gamma_{\vec{u}_1}})$. Also in this case, Eq.~(\ref{eq:pow_engine_max}) only requires a model-dependent maximization over $\vec{u}_1$ and $\vec{u}_2$.

\subsection{Full maximization}
\label{subsec:spectra_opt}
The maximum average power for the refrigerator and the heat engine that we found in Eqs.~(\ref{eq:pow_fridge_max}) and (\ref{eq:pow_engine_max}) has been maximized over the time fractions spent in contact with each bath. However $P^{\rm (max)}_\text{[R]}$ and $P^{\rm (max)}_\text{[E]}$ still need to be maximized over to the experimentally available controls, i.e. over $\vec{u}_1$ and $\vec{u}_2$. Until now, we did not make any assumption on the functional form of $\Gamma_{\vec{u}}$, nor of $H_{\vec{u}}$.

We now assume that the rates $\Gamma_{\vec{u}_i} = \Gamma_{\alpha_i}$ are fixed for each bath (i.e. they do not depend on the value of the control $\vec{u}$, but only on the bath index $\alpha$), and that the control on the Hamiltonians is total (i.e. that we can generate any Hamiltonian). In such case, the maximization of $P^{\rm (max)}_\text{[E]}$ is carried out by maximizing
\begin{equation}
\Tr\left[\left(H_1-H_2\right)\left(\e^{-\beta_1H_1}/Z_1- e^{-\beta_2H_2}/Z_2  \right)\right]
\label{eq:max_engine_spectrum}
\end{equation}
with respect to the choice of the two Hamiltonians $H_1$ and $H_2$.   This maximization has been carried out in Ref.~\cite{Allahverdyan2013}, finding that ${{H}}_1$ and ${{H}}_2$ must be diagonal in the same basis $\ket{\nu}$, and that the spectrum must be given by a non-degenerate ground state, and a $d-1$ degenerate excited state. We therefore have
\begin{equation}
\label{eq:special_spectrum}
{{H}}_i=\sum_{\nu=2}^{d} \varepsilon_i \ket{\nu}\bra{\nu},
\end{equation} 
where $\varepsilon_1$ and $\varepsilon_2$ can be found by maximizing the form taken by \eqref{eq:max_engine_spectrum}, i.e.
\begin{equation}
\dfrac{(\varepsilon_1-\varepsilon_2)(e^{-\beta_1\varepsilon_1}-e^{-\beta_2\varepsilon_2})(d-1)}{(1+(d-1)e^{-\beta_1\varepsilon_1})(1+(d-1)e^{-\beta_2\varepsilon_2})}\ .
\end{equation}

Analogously, it can be shown that the optimization of $P^{\rm (max)}_\text{[R]}$ is carried out by maximizing the numerator of Eq. (\ref{eq:pow_fridge_max}) with respect to the values the Hamiltonian assumes while in contact with the first and the second bath, as we did for Eq. (\ref{eq:max_engine_spectrum}) in the heat engine case.
The optimal Hamiltonians are again of the form of Eq.~\eqref{eq:special_spectrum}, with $\varepsilon_1\rightarrow\infty$ (physically, the hot bath attempts to drive S towards its ground state, to obtain a better cooling), while $\varepsilon_2$ can be obtained as the maximum of the following expression
\begin{equation}
\dfrac{\varepsilon_2 e^{-\beta_2\varepsilon_2}(d-1)}{1+(d-1)e^{-\beta_2\varepsilon_2}}\ .
\label{eq:fridge_max}
\end{equation}

Incidentally, we notice here that the optimality of Hamiltonians with $d-1$ degenerate spectra was found also in the regime opposite to the fast driving, that is in the slow driving, high efficiency regime~\cite{Abiuso2020}.

\subsection{Non-interacting vs many-body qubits}
\label{subsec:many_body}
In this section we compare the maximum GAP of a heat engine and of a refrigerator delivered by $n$ non-interacting qubits [$\text{GAP}_\text{NI}(n)$] driven independently, with the GAP of $n$ interacting qubits [$\text{GAP}_\text{I}(n)$]. In the non-interacting case, we assume that we have full control over the Hamiltonian of the single qubits. In the interacting case, we assume to have full control over the total many-body Hamiltonian of $n$ qubits, i.e. we assume that the control allows us to produce any many-body spectrum. Clearly, this assumption is not expected to hold in specific many-body models. However, this limiting case allows us to find a closed solution which can be seen as an upper bound to the power of any realistic many-body system, and realistic many-body proposals to implement the Hamiltonian in Eq.~(\ref{eq:special_spectrum}) have been put forward \cite{dodds2019}. Furthermore, we work under the assumptions of Secs.~\ref{subsec:spectra_opt}, i.e. we consider rates $\Gamma_{\alpha}$, for $\alpha=1,2$, that only depend on the bath index. 
Under these assumptions, the GAP delivered by a single qubit can be computed as described in Sec.~\ref{subsec:spectra_opt} setting $d=2$. $\text{GAP}_\text{NI}(n)$ will then be equal to $n$ times the power of a single qubit. Instead, $\text{GAP}_\text{I}(n)$ can be computed setting $d=2^n$.

 \begin{figure}[!tb]
	\centering
	\includegraphics[width=0.49\columnwidth]{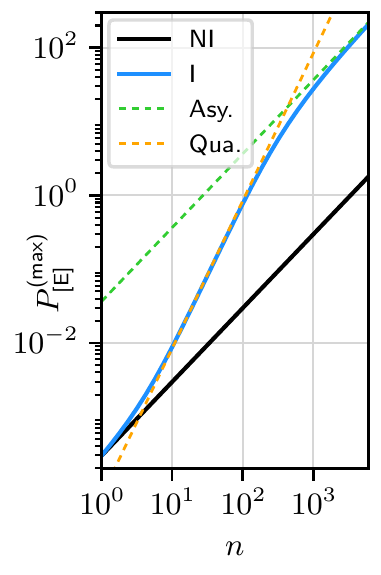}
	\includegraphics[width=0.49\columnwidth]{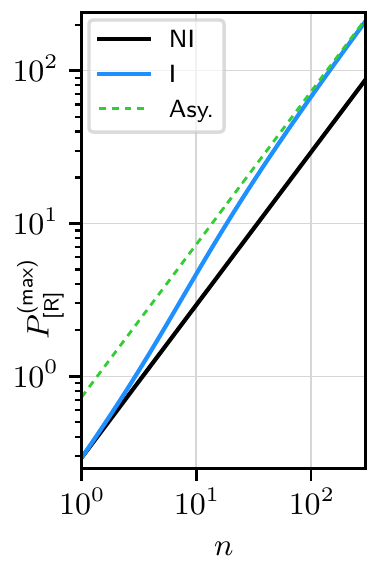}
	\caption{$P_\text{[E]}^\text{(max)}$ (left panel) and $P_\text{[R]}^\text{(max)}$ (right panel), measured in units of $\beta_2^{-1}/(\sqrt{\Gamma_1^{-1}} + \sqrt{\Gamma_2^{-1}})^2$, as a function of the number of qubits $n$ displayed in a log-log plot. The black curve corresponds to the non-interacting case, while the blue curve to the interacting case. The dashed green curve is the analytic asymptotic value we find for $n\to \infty$, see Eq.~(\ref{eq:p_asymp}), while the dashed orange line is a reference quadratic function. The temperatures are chosen such that $\beta_1=0.95\,\beta_2$, and the power scales with the rates and temperatures according to the units it is expressed in. }
	\label{fig:many_qubits}
\end{figure}
In Fig.~\ref{fig:many_qubits} we show the maximum GAP of a heat engine, $P_\text{[E]}^\text{(max)}$ (left panel), and the maximum GAP of a refrigerator, $P_\text{[R]}^\text{(max)}$ (right panel), as a function of the number of qubits $n$ in a log-log plot. The black curve, corresponding to $\text{GAP}_\text{NI}(n)$, is a linear function of $n$. Notably, there is a transient regime, roughly between $10^0$ and $10^2$, where $\text{GAP}_\text{I}(n)$ is superlinear: in particular, $P_\text{[E]}^\text{(max)}$ scales as $n^2$, 
 whereas $P_\text{[R]}^\text{(max)}$ scales as a $n^\delta$, with $1<\delta<2$. However, for large enough $n$ (thermodynamic limit), we see that $\text{GAP}_\text{I}(n)$ is again a linear function of $n$, displaying a finite gap with respect to $\text{GAP}_\text{NI}(n)$. In App.~\ref{app:many_qubits} we prove that the asymptotic behavior is given by
\begin{equation}
\begin{aligned}
	P_\text{[E]}^\text{(max)} \,\, &\overset{n\to\infty}{=}\,\, \ln 2 \,\frac{\beta_1^{-1} - \beta_2^{-1}}{\left(\sqrt{\Gamma_1^{-1}} + \sqrt{\Gamma_2^{-1}}\right)^2} \, n, \\
	P_\text{[R]}^\text{(max)} \,\, &\overset{n\to\infty}{=}\,\, \ln 2 \,\frac{\beta_2^{-1}}{\left(\sqrt{\Gamma_1^{-1}} + \sqrt{\Gamma_2^{-1}}\right)^2} \, n, \\
\end{aligned}
\label{eq:p_asymp}
\end{equation}
which is indeed linear in $n$. 

Remarkably, in the heat engine case, the asymptotic behavior of $P_\text{[E]}^\text{(max)}$ is linear in the temperature difference $\Delta T = (\beta_1^{-1} - \beta_2^{-1})/k_B$.
This is quite surprising: indeed, any finite-size slowly-driven quantum system \cite{Abiuso2020} or any finite-size steady-state thermoelectric heat engine \cite{Benenti2017} delivers a maximum power which, for small $\Delta T$, scales as $\Delta T^2$. Furthermore, also the maximum power of a qubit-based heat engine is proportional to $\Delta T^2$ (see Ref.~\cite{Erdman2019}), yielding $\text{GAP}_\text{NI}(n) = c_0\, n\, (\sqrt{\Gamma_1^{-1}} + \sqrt{\Gamma_2^{-1}})^{-2} \,k_B\Delta T^2/\bar{T}$, where $c_0 \approx 0.11$ and $\bar{T} = (\beta_1^{-1}+\beta_2^{-1})/2$ is the average temperature of the baths. 
This implies that, for small temperature differences and large $n$, 
\begin{equation}
	\text{GAP}_\text{I}(n)/\text{GAP}_\text{NI}(n) \propto 1/(\Delta T/\bar{T}),
\end{equation}
which \textit{diverges} in the limit $\Delta T/\bar{T}\to 0$. This is clear evidence of the advantage of many-body systems over non-interacting systems for the construction of a heat engine. 
 \begin{figure}[!tb]
	\centering
	\includegraphics[width=0.99\columnwidth]{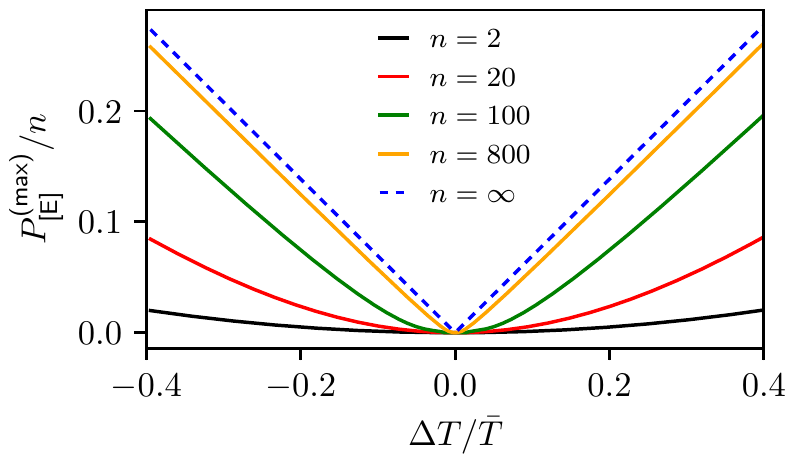}
	\caption{Maximum power-per-qubit in units of $k_B\bar{T}(\sqrt{\Gamma_1^{-1}} + \sqrt{\Gamma_2^{-1}})^{-2}$, as a function of $\Delta T/\bar{T}$, for various values of $n$. The blue dashed lines corresponds to the $|\Delta T|$ scaling predicted in the thermodynamic limit [see Eq.~(\ref{eq:p_asymp})]. The power scales with the chosen parameters according to the units it is expressed in. }
	\label{fig:power_dt}
\end{figure}
Another way to visualize this advantage is provided in Fig.~\ref{fig:power_dt}, where we plot the power-per-qubit, $P_\text{[E]}^\text{(max)}/n$, as a function of $\Delta T/\bar{T}$, maintaining a fixed average temperature $\bar{T}$. Each curve corresponds to a different value of $n$. As we can see, for finite values of $n$, the behaviour of the power-per-qubit is quadratic around $\Delta T=0$. However, as $n$ increases, the power switches to the linear regime for smaller and smaller values of $\Delta T$, approaching the non-analytic $|\Delta T|$ behaviour in the thermodynamic limit (see the dashed line in Fig.~\ref{fig:power_dt}).

Another notable many-body advantage is signalled by the efficiency at maximum power $\eta(P_\text{[E]}^\text{(max)})$,
defined as the ratio between the extracted power and the heat flux provided by the hot bath (both time-averaged over a cycle) when the system is driven at maximum power. As shown in App.~\ref{app:many_qubits}, and in analogy with the findings of Ref.~\cite{Allahverdyan2013}, we find that $\eta(P_\text{[E]}^\text{(max)})$ approaches Carnot's efficiency $\eta_\text{C} = 1- \beta_1/\beta_2$ for large $n$ as
\begin{equation}
    \eta_\text{C} - \eta(P_\text{[E]}^\text{(max)})\,\, \overset{n\to\infty}{=}\,\, \frac{2}{\ln 2}\frac{\beta_1}{\beta_2} \frac{\ln n}{n}.
    \label{eq:eta_max_pow}
\end{equation}
It is therefore possible to asymptotically approach Carnot efficiency at maximum power in this specific model.

In the refrigerator case the comparison between the non-interacting and interacting cases reveals a completely different behavior.
The maximum cooling power can be computed analytically (see App.~\ref{app:many_qubits} for details)
obtaining, in the thermodynamic limit,

\begin{equation}
	\text{GAP}_\text{I}(n)/\text{GAP}_\text{NI}(n) = \ln 2/ W(1/e) \approx 2.49,
	\label{eq:gap_ni_i_fridge}
\end{equation}
where $W(x)$ is the Lambert function.
The previous equation highlights that there is no relevant advantage in using a many-body interacting working fluid over $n$ separate units working in parallel.
Furthermore, the coefficient of performance at maximum power, defined as the ratio between the maximum cooling power and the power provided to the system, turns out to be null both in the interacting and non-interacting cases as a consequence of the fact that $\varepsilon_1\to \infty$ (see App.~\ref{app:many_qubits} for details).

\section{Case study: a Qutrit Heat Engine}
\label{sec:qutrit}
In this section we discuss and apply our optimal strategy to a setup consisting of a qutrit (a three-level system) which can be coupled to two or three thermal baths.  
From our general result, we know that at most $L=d=3$ intervals will be sufficient to maximize the power in the fast driving regime.
 By studying this example, we show that standard models used to describe Fermionic \cite{Nazarov2009} and Bosonic \cite{Breuer2002} baths are optimized simply by $2$ quenches, i.e. through a standard Otto cycle, with a spectrum as the one derived in Sec.~\ref{subsec:spectra_opt}. 
 Then, we explicitly construct an example  
 where the generalized Otto cycle with $3$ quenches outperforms the standard Otto cycle both in the presence of $2$ or $3$ thermal baths. 
 Incidentally, this implies that the power can be enhanced by the presence of more than $2$ heat baths, even if we can only couple the system to one bath at the time. At last we show that - in all cases mentioned above - the power decreases monotonically as we increase the period of the protocol $T$. This is evidence that, in this model, the fast driving regime is indeed the optimal regime to maximize the GAP. 

The Hamiltonian of the system is given by
\begin{equation}
{H}_{\vec{u}(t)} = \epsilon_2(t)\ket{2}\bra{2} +  \epsilon_3(t)\ket{3}\bra{3},
\label{eq:h_qutrit}
\end{equation}
where $\ket{n}$, for $n=1,2,3$, are the three eigenstates with energies $\epsilon_n(t)$. Without loss of generality, we set $\epsilon_1(t)=0$. Our control vector is given by $\vec{u}(t) = (\epsilon_2(t), \epsilon_3(t))$, and we assume that we can couple the system to one bath at the time. 
Following the standard microscopic derivation of the Lindblad master equation (see App. \ref{app:qutrit} for details), the populations $p_n(t)\equiv \langle n| {\rho}(t)\ket{n}$ satisfy \begin{multline}
	{\partial_t}{p}_n(t) = \\
	 \sum_{m\neq n} \left[ -p_n(t)\Gamma_{nm}(\vec{u}(t),\alpha(t)) + p_m(t)\Gamma_{mn}(\vec{u}(t),\alpha(t))  \right],
	 \label{eq:qutrit_pauli}
\end{multline}
where the scalar quantity $\Gamma_{nm}(\vec{u},\alpha)$ represents the transition rate, induced by the bath $\alpha(t)$, from state $\ket{n}$ to state $\ket{m}$. 
Since the baths are assumed to be in equilibrium, the rates satisfy a set of detailed balance conditions 
\begin{equation}
	\Gamma_{nm}(\vec{u},\alpha) = e^{\beta_\alpha (\epsilon_n - \epsilon_m)} \Gamma_{mn}(\vec{u},\alpha),
	\label{eq:dbe}
\end{equation}
which fix half of the rates. With this notation, the GAP of a heat engine, i.e. Eq.~(\ref{eq:pa1}) with $c_\alpha=1$, is given by (see App. \ref{app:qutrit} for details)
\begin{equation}
	P_{[E]}[\vec{u}]  = \sum_n \epsilon_n(t) {\partial_t}{p}_n(t).
	\label{eq:pe_qutrit}
\end{equation}

When the baths are given by a continuum of free Fermionic (F) or Bosonic (B) particles, and the Hamiltonian describing the system-bath coupling is quadratic, the rates are given by \cite{Nazarov2009, Breuer2002}
\begin{equation}
\begin{aligned}
	\Gamma_{nm}^\text{(F)}(\vec{u},\alpha) &= \gamma(\alpha) f[\beta_\alpha(\epsilon_m-\epsilon_n)], \\
	\Gamma_{nm}^\text{(B)}(\vec{u},\alpha) &= \gamma(\alpha) (\epsilon_m-\epsilon_n) n[\beta_\alpha(\epsilon_m-\epsilon_n)], \\
\end{aligned}
\label{eq:fermi_bose_rates}
\end{equation}
where $f(x)=(e^x +1)^{-1}$ and $n(x)=(e^x -1)^{-1}$ are respectively the Fermi and Bose distributions. Notice that, in Eq.~(\ref{eq:fermi_bose_rates}), we are assuming the spectral density to be flat in the Fermionic case, and Ohmic in the Bosonic case. Physically, these models may describe respectively electronic leads coupled to a quantum dot \cite{Esposito2009,Nazarov2009, Esposito2010,Koski2014,Erdman2017,Josefsson2018, Prete2019} or photonic baths coupled to an (artificial) atom \cite{Breuer2002, Geva1992, Alicki1979, Ronzani2018, Senior2020}. 

Applying the results of the previous sections to these models, as detailed in App.~\ref{app:qutrit}, we have that the optimal protocol maximizing Eq.~(\ref{eq:pe_qutrit}) in the fast driving regime is an Otto cycle with $L=3$. This last is completely specified by $8$ parameters: two time fractions $\mu_i$ (since $\sum_i \mu_i=1$) and $6$ control values $\epsilon_n^{(i)}$, where we defined  $\epsilon_n^{(i)}$ as the value of $\epsilon_n(t)$ during the time interval $i=1,2,3$. Notice that we only optimize over $\epsilon_n^{(i)}$ for $n=2,3$ since $\epsilon_1(t)=0$. Assuming that we have two thermal baths at inverse temperatures $\beta_2=2\beta_1$, we study all the possible configurations of the values of the temperature in each one of the time intervals.

 \begin{figure}[!t]
	\centering
	\includegraphics[width=1\columnwidth]{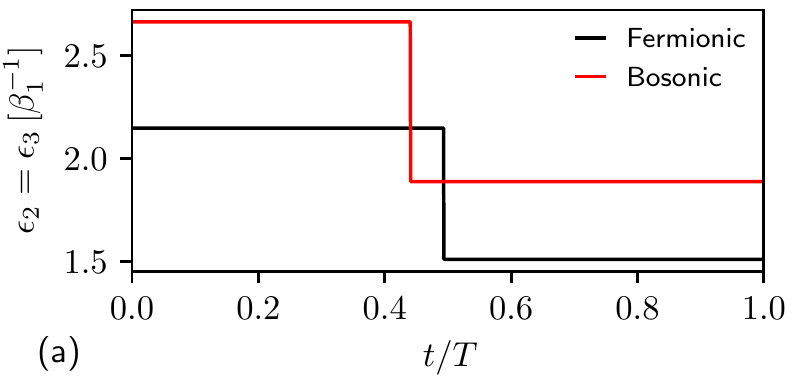}
	\includegraphics[width=1\columnwidth]{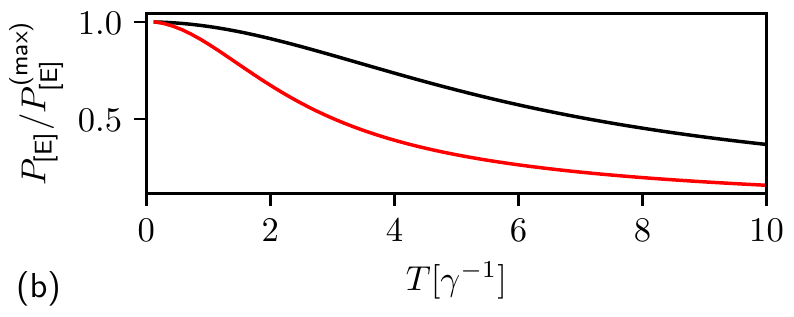}
	\caption{(a) The optimal protocol, described by $\epsilon_1(t)=0$ and $\epsilon_2(t)=\epsilon_3(t)$, as a function of time $t$ normalized to the protocol period $T$ for the Fermionic and Bosonic models. The system is first coupled to bath $\alpha=1$, then to $\alpha=2$ after the quench. (b) Average power, normalized to the peak value, as a function of the period $T$. In both panel the parameters are $\beta_2=2\beta_1$ and $\gamma(\alpha)=\gamma$, where $\gamma$ is a fixed timescale. Given these parameters, the protocol scales as the unit it is expressed in, while the ratio $P_{[\text{E}]}/P^{(\text{max})}_{[\text{E}]}$ remains constant.}
	\label{fig:qutrit_fermionic_bosonic}
\end{figure}

Carrying out this last stage of the optimization numerically, we find that both the Bosonic and the Fermionic models are optimized by a standard Otto cycle with only $2$ quenches. Furthermore, the maximum power is achieved when $\epsilon_2^{(i)} = \epsilon_3^{(i)}$, which is exactly the energy spectrum which we proved to be optimal in a simpler relaxation model, see Sec.~\ref{subsec:spectra_opt}. In Fig.~\ref{fig:qutrit_fermionic_bosonic}(a), we plot the optimal protocol, described by $\epsilon_1(t)=0$ and $\epsilon_2(t)=\epsilon_3(t)$, as a function of time, while in Fig.~\ref{fig:qutrit_fermionic_bosonic}(b) we plot a finite-time numerical calculation of the average extracted power $P_\text{[E]}$ as a function of the protocol duration $T$, while maintaining the time fractions $\mu_i$ constant. Interestingly, we notice that in both models the power decreases monotonically with $T$, providing us with evidence that - in this case - the fast driving regime may be optimal to maximize power extraction. We also notice that the derivative of $P_\text{[E]}$ respect to $T$, at $T=0$, is null, hinting that our fast driving results may hold up to second order in $\gamma T$. Furthermore, we verified numerically that a large fraction of the maximum power can still be extracted even when the driving is slower than the characteristic rate $\gamma$, showing that our upper bound is ``robust'' to finite-time corrections.

We now show that there are cases in which a generalized Otto cycle with $3$ quenches can outperform a standard Otto cycle. Let us consider a case where the rates $\Gamma_{nm}(\vec{u},\alpha)$ are vanishingly small for all controls $\vec{u}$, except for a set of discrete points. Physically, this could be implemented through peaked density of states in the baths. Within this assumptions,  we can design $3$ thermal baths at inverse temperatures $\beta_\alpha$ (for $\alpha=1,2,3$), such that each one induces a non-null rate only when the controls $\epsilon_2(t)$ and $\epsilon_3(t)$ take the values $\tilde{\epsilon}^{(\alpha)}_2$ and $\tilde{\epsilon}^{(\alpha)}_3$, respectively. Mathematically, we can describe this scenario defining
\begin{equation}
	\Gamma_{nm}(\vec{u},\alpha) = \gamma_{nm}(\alpha)\,\chi(\vec{u},\alpha),
\end{equation}
where $\gamma_{nm}(\alpha)$ are constants, and $\chi(\vec{u},\alpha)$ is an indicator function equal to one for $\epsilon_2(t) = \tilde{\epsilon}^{(\alpha)}_2$ and $\epsilon_3(t) = \tilde{\epsilon}^{(\alpha)}_3$ and zero otherwise.
In this scenario, the optimal values of $\epsilon^{(i)}_n$ are simply given by $\tilde{\epsilon}^{(i)}_n$, the only parameters over which we must optimize being the time fractions $\mu_i$, which we express as
\begin{equation}
\begin{aligned}
	\mu_1 &= x,\quad & \mu_2 &= (1-x)y,\quad & \mu_3 &= (1-x)(1-y),
\end{aligned}
\end{equation}
where $x\in [0,1]$ and $y\in [0,1]$ cover all possibilities. This parameterization is such that one of the three time fractions $\mu_i$ is null if and only if $x$ and/or $y$ takes the values $0$ or $1$. 
 \begin{figure}[!tb]
	\centering
	\includegraphics[width=1\columnwidth]{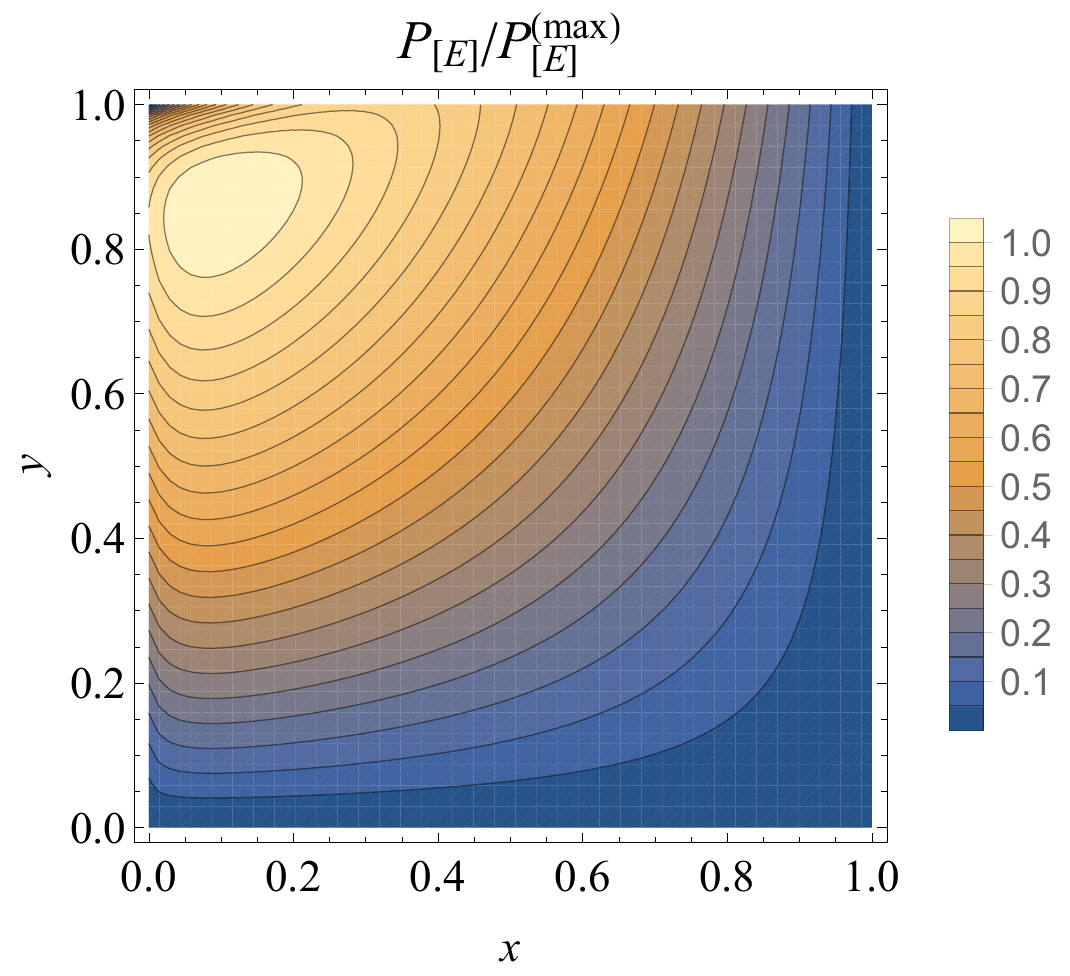}
	\caption{Contour plot of the average power, normalized to the peak value $P_\text{[E]}^\text{(max)}$, as a function of $x$ and $y$. The parameters are: $\beta_1 = 8.12 \beta_2$, $\beta_3 = 7.81 \beta_2$, $\epsilon_2^{(1)}=1.85 \beta_2^{-1}$, $\epsilon_3^{(1)}=1.56 \beta_2^{-1}$, $\epsilon_2^{(2)}=10.07 \beta_2^{-1}$, $\epsilon_3^{(2)}=9.58 \beta_2^{-1}$, $\epsilon_2^{(3)}=1.75 \beta_2^{-1}$, $\epsilon_3^{(3)}=8.12 \beta_2^{-1}$, $\gamma_{12}(1) = \gamma$, $\gamma_{13}(1) = 1.21 \gamma$, $\gamma_{23}(1) = 2.28 \gamma$, $\gamma_{12}(2) = 9.45\gamma$, $\gamma_{13}(2) = 2.53\gamma$, $\gamma_{23}(2) = 5.26\gamma$, $\gamma_{12}(3) = 5.9\gamma$, $\gamma_{13}(3) =1.4 \gamma$, $\gamma_{23}(3) = 6.22 \gamma$.}
	\label{fig:3_quench}
\end{figure}
In Fig.~\ref{fig:3_quench} we show a contour-plot of the average power $P_\text{[E]}$ as a functions of $x$ and $y$ (see caption for the parameters used). As we can see, the maximum power does not occur on the sides of the box: this implies that all optimal $\mu_i$ are finite, so the generalized Otto cycle with three finite intervals outperforms the standard Otto cycle. Numerically, we find that the optimal power occurs at $x\approx 0.092$ and $y\approx 0.86$. This result also proves that the power can be enhanced by using three thermal baths; this is in contrast with the optimization of the efficiency, which only requires coupling to the hottest and the coldest bath. Physically, this is due to the fact that, even if the third bath has a temperature between the coldest and the hottest, it may have a thermalization rate which is higher than the other baths, thus speeding up the heat exchange.
We want to remark that there is a range of the parameters space in which $3$ quenches still outperform $2$ quenches even with $\beta_2=\beta_3$ that is, when only two temperatures are available.

At last, we computed the exact finite-time average power $P_\text{[E]}$ as a function of $T$, performing the optimal protocol found through the maximization in Fig.~\ref{fig:3_quench}. Notably, as in the Fermionic and Bosonic models, we find that the power decreases monotonically as $T$ increases, with the same qualitative behavior as in Fig.~\ref{fig:qutrit_fermionic_bosonic}(b).

\section{Non-commuting case}
\label{sec:non-commute}

Throughout the paper, we assumed the dynamics of the working substance to be described by a Markovian master equation derived with the GKSL method.
This derivation can be hindered by non adiabatic effects 
when $[H_{\vec{u}_1},H_{\vec{u}_2}]\neq 0$.
In this case, the dissipators in  (\ref{eq:master_eq_rho}) can depend on the control vector $\vec{u}(t)$ in a non local way, involving, for instance, a dependence on the speed of modulation of $\vec{u}$ \cite{Grifoni1998, Dann2020}.
Such a non local dependence invalidates the proof done in sec.~\ref{sec:fast_driving}.
However, it can still be interesting to investigate how our results can be generalized to the non commuting case, supposing by hypothesis that a time local dynamical generator, like the one in Eq. (\ref{eq:master_eq_rho}), holds also in the fast driving regime.
From a formal point of view, this is a well defined problem and can give us some interesting insights on the physics of the optimal solutions.

In a general scenario in which $[H_{\vec{u}_1},H_{\vec{u}_2}]\neq 0$ there is no reason to suppose, as done in sec. \ref{sec:fast_driving}, that the GAP only depends on the diagonal component of $\rho(t)$, since the dynamical equations of the populations and of the coherences are not decoupled.
The limit cycle of such an evolution is not necessarily restricted to the sole populations, and we have to replace the factor $D = d-1$ in Sec. 
\ref{sec:average_partition} as well as in appendix \ref{app:infinitesimal} with a more general $D= d^2-1$, since the curve parametrizing the state of the system in the cut-and-choose procedure needs now $d(d-1)$ more parameters, representing the coherences, to be fully described.

Under the usual assumptions of irreducibility and adjoint stability, the same arguments of sec. \ref{sec:fast_driving} can be applied to conclude that
 any GAP is maximized by a generalized Otto cycle with maximum number of steps equal to $L= d^2$. We thus observe that:

\begin{enumerate}
    \item The different bound on the number of quenches in the commuting ($L\leq d$) and non-commuting ($L \leq d^2$) cases is a clear signature of how the coherent nature of a dynamical evolution can strongly affect the form of the protocols maximizing the GAP. In this spirit, we can analyze any given maximum power solution in the fast driving regime and recognize {\it a posteriori} if the coherences play a relevant role in the power maximization by checking if
$L> d$.
    \item The generators considered in this section are the most general time local, parameter dependent generators of a Markovian quantum evolution.
    Even if the derivation of the GKSL master equation is not guaranteed to hold in the fast driving regime, such generators may be derived by other means~\footnote{
    See for instance the Caldeira-Leggett master equation, that do not rely on specific assumptions on the time scales of the driving to be derived.}.
    In this sense, the considerations done here go beyond the dissipative dynamics described by a GKSL model.
\end{enumerate}

\section{Conclusions}
\label{sec:conclusions}
In this paper we exhaustively discussed the optimization of thermal machines in the fast driving regime for commuting Hamiltonians.
We proved in full generality that the optimal protocols are universally given by generalized Otto cycles, which are composed by a certain number $L$ of infinitesimal time intervals where the control is fixed. We then bounded $L$ from above in terms of the dimension of the Hilbert space of the working fluid. The proof holds regardless of the specific choice of the control-dependent dissipators, of possible constraints on the control parameters, and regardless of the specific form of the Hamiltonian of the working fluid.

We showed that the standard fast Otto cycle (characterized by $L=2$) is optimal in a vast class of systems. Assuming full control over the system, we explicitly found the optimal driving strategy, which involves producing highly degenerate states, revealing an interesting connection with the results of \cite{Allahverdyan2013} and \cite{Abiuso2020}. 
We then applied this optimal strategy to compare the performance of a refrigerator and of a heat engine based on $n$ interacting and non-interacting qubits. In the refrigerator case, we found that the  non-interacting qubits perform almost as well as the interacting ones; it is therefore reasonable to consider constructing a refrigerator operating in parallel many simple independent units. Conversely, in the heat engine case we found a many-body advantage resulting in the enhancement of both the maximum power, and of the efficiency at maximum power, which approaches Carnot efficiency in the limit of many qubits.

Besides their theoretical relevance, these results lead to a great simplification in the optimization of thermodynamic problems from a practical point of view, due to the intrinsic simplicity of the 
generalized Otto cycle.
This simplification can be exploited both for analytical and numerical treatments, as we explicitly showed studying a  qutrit-based heat engine. In this setup, we analyzed typical configurations, such as fermionic and bosonic baths, and we found a specific form of the dissipators  such that the optimal protocol consists of coupling the system to three baths at different temperatures.
This result marks a difference with the maximization of the efficiency, that always requires only two thermal sources, and shows that the bound on the number of intervals derived in the first part of the paper is actually tight in this case.

This work unlocks the possibility of analytically and/or numerically optimizing the performance of many quantum thermal machines. As future directions, it is interesting to assess the role of coherence in the non-commuting case, and to understand for which classes of systems the fast driving regime is optimal for power extraction. Furthermore, by providing strict bounds on optimal protocols, our results can be used as benchmarks to assess if effects beyond the Markovian regime and weak coupling approximation can indeed enhance, or decrease, the performance of thermal machines. By highlighting the importance of many-body interactions for the performance of a heat engine regime, a future venue would be to identify and study realistic systems which display the $\Delta T$ scaling of the maximum power in the thermodynamic limit. At last, it seems natural to investigate the properties of the fast-driving regime respect to other thermodynamic figures of merit, such as the efficiency at maximum power, or work fluctuations.

\section{Acknowledgments} 
We would like to thank Mart\'{i} Perarnau Llobet for useful conversations and for organizing the ``Quantum Thermodynamics for Young Scientists'' conference together with Philipp Strasberg. We acknowledge fruitful conversations with Rosario Fazio and Fabio Taddei. V.G. acknowledges support from PRIN 2017 “Taming complexity with quantum strategies”. P.A.  is  supported  by “la Caixa” Foundation (ID 100010434, fellowship code LCF/BQ/DI19/11730023), Spanish MINECO (QIBEQI FIS2016-80773-P, Severo Ochoa SEV-2015-0522), Fundacio Cellex, Generalitat de Catalunya (SGR 1381 and CERCA Programme).
V.C. is  funded  by  the National Research Fund of Luxembourg in the frame of project QUTHERM C18/MS/12704391.

This article is dedicated to the memory of Federico Tonielli.

\clearpage

\begin{appendix}
\begin{widetext}

\section{Projected form of the ME} 
\label{app:projection}
Using  projection techniques in this section we show  how one
 can cast the master equation~(\ref{eq:master_eq_rho}) in the more convenient form~(\ref{eq:dyn2}). 
 For this purpose we find it useful to first recall some structural properties of GKSL generators which hold true in the finite
  dimensional case we are analyzing in the present work. 
 In particular in Secs.~\ref{SEC:ERGO} and \ref{app:irr_gksl} we shall introduce the notions of ergodicity, mixing, irreducibility, and adjoint-stability.
After that we proceed with the derivation of Eq.~(\ref{eq:dyn2}) in Sec.~\ref{DERIVATIONAPPA}.

\subsection{Ergodic, Mixing, Irreducible and adjoint-stable GKSL generators} \label{SEC:ERGO} 

Let S be a quantum system of finite dimension $d$.
To fix the notation we indicate with  ${\mathfrak{L}}_{\text{S}}$ 
the  $d^2$-dimensional vector space  of linear operators on S
and define ${\mathfrak{S}}_{\text{S}}$ and ${\mathfrak{L}}^{0}_{\text{S}}$  its subsets formed respectively by the density  and zero-trace operators of the model, i.e. 
\begin{eqnarray} 
{\mathfrak{S}}_{\text{S}}  \equiv \left\{ \rho \in {\mathfrak{L}}_{\text{S}}| \mbox{Tr}[\rho]=1\;, \rho\geq 0 \right\}\;, 
\qquad \qquad
{\mathfrak{L}}^{0}_{\text{S}}\equiv \left\{ \Theta \in {\mathfrak{L}}_{\text{S}}| \mbox{Tr}[\Theta]=0\right\}\;.
\end{eqnarray} 
The 
the latter forms a ($d^2-1$)-dimensional vector subspace of ${\mathfrak{L}}_{\text{S}}$ for which we can identify a projector introducing the super-operator 
\begin{eqnarray}
\mathcal{Q}[\cdots] &\equiv& \text{Id}[\cdots]  - \mbox{Tr}[\cdots]\frac{\mathbb{I}}{d}\;,  
\end{eqnarray} 
and its orthogonal complement
$\mathcal{P} \equiv \text{Id} - \mathcal{Q}$, $\text{Id}$ being the identity channel (notice in fact that 
using ``$\circ$'' to represent the composition of super-operators we have 
$\mathcal{Q}\circ 
\mathcal{Q} = \mathcal{Q}$, $\mathcal{P}\circ 
\mathcal{P} = \mathcal{P}$, $\mathcal{P}\circ 
\mathcal{Q} = \mathcal{Q}\circ 
\mathcal{P}=0$, and that  
$\mbox{Tr}[\mathcal{Q}[\Theta]]=0$ with $\mathcal{Q}[\Theta]=\Theta$ iff $\Theta \in {\mathfrak{L}}^{0}_{\text{S}}$).

  Consider next a GKSL
 generator $\cal L$ for a generic  time-independent master equation 
 \begin{eqnarray} \label{equation1}
 \partial_t \rho(t) = \mathcal{L} [\rho(t)] \;, \end{eqnarray} 
 for the density matrices of S.   By general properties of the theory we know that $\cal L$ is a super-operator 
on ${\mathfrak{L}}_{\text{S}}$ which can always be cast in the standard form 
\begin{eqnarray}
{\cal L}[\cdots] &=& -i[H, \cdots ] + {\cal D}[\cdots]\;, \\
	{\cal D}[ \cdots] &=& \sum_j \left( A_j[\cdots]  A_j^\dagger -\frac{A_j^\dagger A_j[\cdots] +[\cdots] A_j^\dagger A_j}{2}
	  \right), \label{Lind}
\end{eqnarray}
where $H$ is a hermitian operator identifying the Hamiltonian of the system, and ${\cal D}$ is a purely dissipator component written in terms of  the (Lindblad) operators~$A_j\neq 0$.
It is also a well know fact that $\mathcal{L}$ transforms any $\Theta\in {\mathfrak{L}}_{\text{S}}$ 
 into an element of  traceless subset ${\mathfrak{L}}^{0}_{\text{S}}$ (i.e. $\mbox{Tr}[{\cal L}[ \Theta]] =0$),
 which   formally translates into the following identity 
 \begin{eqnarray} \label{dffdtrace} 
\mathcal{Q}  \circ {\cal L} =  {\cal L}\;,
\end{eqnarray} 
and that it
 admits always at least a fix-point state ${\rho}^{\text{(eq)}}\in {\mathfrak{S}}_{\text{S}} $, i.e. a density matrix of S which is an eigen-operator 
 of  $\mathcal{L}$ associated with  the eigenvalue zero, 
  \begin{eqnarray} \label{dffd} 
{\cal L}[{\rho}^{\text{(eq)}}]= 0  \;.
\end{eqnarray} 
Thanks to the above properties we can hence observe that for all density matrices $\rho$, we have
\begin{eqnarray} \label{IDENTITY1} 
{\cal L} [\rho] = (\mathcal{Q}  \circ {\cal L})[\rho]=(\mathcal{Q}  \circ {\cal L})[\rho -{\rho}^{\text{(eq)}}]= 
(\mathcal{Q}  \circ {\cal L}\circ \mathcal{Q} )[\rho -{\rho}^{\text{(eq)}}]=(\mathcal{Q}  \circ {\cal L}\circ \mathcal{Q} )[\tilde{\rho} -\tilde{\rho}^{\text{(eq)}}]\;,
\end{eqnarray} 
where in the third passage we use the fact that $\rho -{\rho}^{\text{(eq)}}$ has trace zero, while in the final one we adopt the short hand notation  $\tilde{\Theta}$  to indicate the projected component of 
$\Theta$ on $\mathfrak{L}^{0}_{\text{S}}$, i.e. 
\begin{eqnarray} \label{SIMPLIFQ} 
\tilde{\Theta} \equiv {\cal Q}[\Theta]\;. 
\end{eqnarray} 
Notice finally that from $\rho(t) = ({\cal Q} + {\cal P})[\rho(t)] = \tilde{\rho}(t) + \frac{\mathbb{I}}{d}$ follows that 
$\partial_t \rho(t) = \partial_t \tilde{\rho}(t)$. Thus using this and~(\ref{IDENTITY1}) evaluated for $\rho=\rho(t)$, we can hence 
conclude that an equivalent way to express Eq.~(\ref{equation1}) is  
\begin{eqnarray} \label{equation1eq}
 \partial_t \tilde{\rho}(t) = \mathcal{G} [\tilde{\rho}^{\text{(eq)}}-\tilde{\rho}(t)] \;, \end{eqnarray} 
where 
 \begin{eqnarray}{\cal G} \equiv - \mathcal{Q}\circ {\cal L}  \circ \mathcal{Q} =- {\cal L}  \circ \mathcal{Q}\;,\label{oldGAMMA}
\end{eqnarray} 
is (minus) the restriction of $\cal L$ on  $\mathfrak{L}^{0}_{\text{S}}$.
\\

Equation~(\ref{equation1eq}) is valid for all the finite-dimensional GKSL processes, but it becomes particularly handy when specified
under ergodicity assumptions~\cite{Burgarth2013}, i.e. for those ${\cal L}$   for which the fix-point state 
${\rho}^{\text{(eq)}}$ introduced in Eq.~(\ref{dffd}) constitute the unique
 eigenvectors with zero eigenvalue. 
\\

{\bf Definition:}  {\it The generator $\mathcal{L}$ is said to be {\bf ergodic} 
if } ${\rho}^{\text{(eq)}} \in  {\mathfrak{S}}_{\text{S}} $  {\it exists such that}
\begin{eqnarray} \label{dffd1} 
{\cal L}[{\Theta}]= 0  \quad \Longleftrightarrow \quad {\Theta} =\lambda {\rho}^{\text{(eq)}}\;,
\end{eqnarray} 
 {\it where $\lambda$ is an arbitrary complex constant.} 
 \\
 
For our purposes, the main consequence of the above definition is that  for an {ergodic} GKSL generator $\mathcal{L}$ the corresponding
restriction $\cal G$ defined in Eq.~(\ref{oldGAMMA})  is 
invertible  when acting on the elements of the $(d^2-1)$-dimensional linear subspace $\mathfrak{L}^{0}_{\text{S}}$.
Indeed using the fact  that ${\rho}^{\text{(eq)}}$  has trace 1, 
 we can conclude that under
 ergodic assumption~(\ref{dffd1}) it holds 
\begin{eqnarray} \label{ffd1} 
{\cal G} [\Theta] = 0 \quad \Longleftrightarrow  \quad  {\cal Q}[\Theta] =0\;,
\end{eqnarray} 
or equivalently that 
${\cal G}$ has no zero eigenvalue on   $\mathfrak{L}^{0}_{\text{S}}$.

The ergodicity property~(\ref{dffd1}) has been extensively studied in several works. In particular a
 necessary and sufficient condition for ${\cal L}$ to be ergodic can be found e.g. in Ref.~\cite{Burgarth2013} where it has been also shown  that
this property is very common on the set of the GKSL generators (the non-ergodic examples being indeed a set of zero measure). 
Interestingly enough it turns out that at least for the finite dimensional case we are studying here, Eq.~(\ref{dffd1}) is equivalent to asking that the associated  ME
should induce a purely mixing evolution which 
asymptotically sends all input states ${\rho} \in \mathfrak{S}_\text{S}$  of the system into ${\rho}^{\text{(eq)}}$, i.e. 
\begin{eqnarray} 
\lim_{ t \rightarrow \infty}  \label{imposing1} 
\| {\rho}(t) - {\rho}^{\text{(eq)}} \|_1 =0 \;,
\end{eqnarray} 
where $\rho(t) = e^{t \mathcal{L}}[\rho]$ is the completely positive evolution obtained by integrating~(\ref{equation1})  and $\| \cdots \|_1$ is the 
trace norm 
(the fact that (\ref{imposing1}) implies (\ref{dffd1})  is relatively easy to verify, while an explicit proof of the opposite implication can be
found in Refs.~\cite{Holevo2001,Alicki2007,Schirmer2010,Baumgartner2012}). 

The main drawback of the ergodic property is that~(\ref{dffd1}) does not behave well under  summation of the GKSL generators, i.e. the sum of ergodic generators is not necessarily ergodic
(see~\cite{Burgarth2013} for an explicit counterexample). Nonetheless, a slightly stronger version of the ergodicity notion does not suffer from this limitation.
This is the set of GKSL generators $\mathcal{L}$ which are {\bf irreducible}  and {\bf adjoint-stable}~\cite{Spohn1980,Menczel2019}: 
\\

{\bf Definition:}  {\it Given a GKSL generator ${\cal L}$ and} ${\cal A}\equiv \mbox{Span}\{ A_i\}$ 
{\it the set spanned  by its Lindblad operators, we say that}
${\cal L}$ is {\bf irreducible} if $[A, B] =0$ for all $A\in {\cal A}$ 
implies that $B = \lambda \openone$ for some complex number  $\lambda$,
and that 
$\mathcal{L}$ is {\bf adjoint-stable} if $A\in {\cal A}$
implies  $A^\dagger \in {\cal A}$.
 \\

First of all it is worth noticing that  both these two properties only involve the dissipative component ${\cal D}$  of ${\cal L}$ (indeed they are independent of the system Hamiltonian $H$).
Secondly, as discussed in Refs.~\cite{Spohn1980,Menczel2019} it follows that all  $\mathcal{L}$ which are {\bf irreducible} and  {\bf adjoint-stable} induce 
dynamical evolutions which are mixing (i.e. obey to Eq.~(\ref{imposing1}) with ${\rho}^{\text{(eq)}}$ being identified with the steady state solution of the model) and
hence, via the above mentioned equivalence, ergodic, i.e.  
\begin{eqnarray} \label{IMPO111} 
\mbox{$\mathcal{L}$ {\bf irreducible} and  {\bf adjoint-stable}} \Longrightarrow \mbox{$\mathcal{L}$ {\bf ergodic}.}
\end{eqnarray} 
Most importantly it also follows that, at variance with the ergodic set,  the set of irreducible and adjoint-stable GKSL generators is closed under
summation (in particular they form a convex set): more specifically  given ${\cal L}$
 {irreducible} and  {adjoint-stable}, and 
 ${\cal L}'$ adjoint-stable but not necessarily  irreducible,   their sum is 
 {irreducible} and  {adjoint-stable}, i.e. 
\begin{eqnarray} \label{SUMRULE} 
\mbox{$\mathcal{L}$ {\bf irreducible} and  {\bf adjoint-stable}}, \mbox{$\mathcal{L}'$ {\bf adjoint-stable}} \Longrightarrow 
\mbox{$\mathcal{L}+\mathcal{L}'$ {\bf irreducible} and  {\bf adjoint-stable}.} \end{eqnarray} 

We now focus on a special subset of ergodic GKSL generators ${\cal L}$ which provide a rather general 
description  of 
thermalization events, see e.g.~\cite{Breuer2002}.
\\

{\bf Definition:}  {\it Given $\beta\geq 0$, a generator GKSL $\mathcal{L}$ 
is said to be  {\bf thermalizing}  
 if it is adjoint-stable and ergodic with fixed point provided by the Gibbs density matrix}
 \begin{eqnarray} \label{GIBBS} 
\rho^\text{(eq)}_{\beta} \equiv \exp[ -\beta H]/Z_\beta\;, \qquad Z_\beta\equiv \mbox{Tr}[\exp[ -\beta H]]\;. 
\end{eqnarray} 

Notice that requiring adjoint-stability for a thermalizing map is in agreement with the underlying open system derivation of the master equation.
Indeed, if this last is derived from a microscopic model in which the system is weakly coupled to a thermal bath of inverse temperature
$\beta$, the adjoint stability of the GKSL generator can be proven
using the Kubo-Martin-Schwinger relations for the bath correlation functions \cite{Breuer2002}.

 We now claim that a thermalizing generator satisfies also a
 weak notion of irreducibility. To begin we notice that in Eq.~(\ref{GIBBS})  the parameter $\beta$ plays the role of an inverse temperature and that, for all finite values of such quantity, the density matrix $\rho^\text{(eq)}_{\beta}$ is a full rank state (for $\beta\rightarrow \infty$, i.e. as the temperature drops to zero,
  this property is not longer guaranteed as $\rho^\text{(eq)}_{\beta}$  converges to the ground state of $H$). 
  Accordingly we can invoke Theorem 5.2 of~\cite{Spohn1980} to claim that for thermalizing processes the linear set
  ${\cal A}_H \equiv \mbox{Span}\{ A_i, H\}$ spanned by the Lindblad operators $A_i$ and by the Hamiltonian $H$ is irreducible, i.e. that the following implication holds
  \begin{eqnarray} \label{WIRR} 
  [B, A] =0 \qquad \forall A\in {\cal A}_H\;, \qquad \Longrightarrow B = \lambda \openone\;,
  \end{eqnarray} 
  with $\lambda$ generic complex constant. 
We will refer to the condition (\ref{WIRR}) as to {\bf weak irreducibility},
since it is less demanding than the standard irreducibility, that can be recovered 
with some assumptions on the nature of the Lindblad operators. 

\subsection{Irreducibility for physical GKSL generators}
\label{app:irr_gksl}
Within the assumption of a Thermalizing GKSL generator, let us suppose the Lindblad operators 
to be represented by jump operators, i.e. of the
form $\sqrt{\gamma_{E_j \rightarrow E_{i}}} |E_i\rangle\langle E_j|$ where $|E_i\rangle$ denotes an eigenvector of the system Hamiltonian,
whose levels are assumed to be non-degenerate.

In this case we have, for every operator $B = \sum_{i',j'} m_{i',j'} |E_{i'}\rangle\langle E_{j'}|$ with $0\leq i',j'\leq D$:

\begin{equation}
\big[B,  \sqrt{\gamma_{E_j \rightarrow E_{i}}} |E_i\rangle \langle E_j|\big] =\sqrt{\gamma_{E_j \rightarrow E_{i}}} \big( \sum_{i',j'} 
 m_{i',i} |E_{i'}\rangle\langle E_j| +  m_{j,j'} |E_{i}\rangle\langle E_{j'}| \big). \label{comm}
\end{equation}
 
Using the equation above, we have

\begin{equation}
  \big[B,  \sqrt{\gamma_{E_j \rightarrow E_{i}}} |E_i\rangle \langle E_j| \big] = 0 \quad \leftrightarrow
  \quad m_{i,i}=m_{j,j}, \; \; m_{i',i}= m_{j,j'}=0 \; \; \forall i'\neq i, \; j' \neq j. \label{condcomm}
\end{equation}

Since the condition $[B,H] =0$ does not constrain the diagonal of $B$, the only way to  
obtain $B = \lambda \openone$ as required by Eq. (\ref{WIRR}) and fulfil weak irreducibility is 
that the set
$\cal{A}_{H}$ contains jumps connecting all the energy levels, i.e. for every $i$ there is at least one Lindblad 
operator $  \sqrt{\gamma_{E_i \rightarrow E_{k}}} |E_k\rangle \langle E_i| $ connecting 
$|E_i\rangle$ with some other level $|E_k\rangle$.
In this way, the first of the two conditions in the r.h.s. of (\ref{condcomm}) imposes that all the diagonal elements of $B$
are equal.
In addition, using the second condition in the r.h.s. of (\ref{condcomm}) we can set all the off diagonal elements to $0$ eventually 
obtaining $B= \lambda I$.
Since the latter has been proven without imposing $[H,B]=0$, we just proved that by choosing the jump operators as Lindblad operators we have 
that weak irriducibility implies {\bf irreducibility}.

More in general, a Lindblad operator is the sum of jump operators connecting couples of levels with the same energy 
difference \cite{Breuer2002}, $A(\omega) = \sum_{l,l'} f_{l,l'} 
|E_l\rangle \langle E_{l'}|$ where $E_l - E_{l'} = \omega$ $\forall l,l'$.

We analyze the condition (\ref{condcomm}) in the 
case of a Lindblad operator composed by two jumps, that is 

\begin{equation} [B,f_{i_1,j_1} |E_{i_1}\rangle \langle E_{j_1}| + f_{i_2,j_2} |E_{i_2}\rangle \langle E_{j_2}|] = 0,   \label{canc2} \end{equation}

we obtain that this last is equivalent to require the r.h.s. of the (\ref{condcomm}) for 
both $i_1,j_1$ and $i_2,j_2$, with the exception that the four off diagonal elements
$ m_{i_1,i_2}$, $m_{j_1,j_2}$, $m_{i_2,i_1}$, $m_{j_2,j_1}$ are not necessarily $0$, but satisfy 
the following linear system

\bg  f_{i_2,j_2} m_{i_1,i_2} - f_{i_1,j_1} m_{j_1,j_2} =0, \\
 f_{i_1,j_1} m_{i_2,i_1} - f_{i_2,j_2} m_{j_2,j_1} =0 . \label{systemapp} \eg
 
Using the adjoint stability property, the (\ref{systemapp}) preserves its validity 
when replacing the elements of $B$ with the ones of $B^{\dagger}$, thus 

\bg  f_{i_2,j_2}^* m_{i_2,i_1} - f_{i_1,j_1}^* m_{j_2,j_1} =0, \\
 f_{i_1,j_1}^* m_{i_1,i_2} - f_{i_2,j_2}^* m_{j_1,j_2} =0 . \label{systemapp2} \eg
 
The solution of Eqs. (\ref{systemapp}) and (\ref{systemapp2}) is the null vector provided that 
$|f_{i_2,j_2}|^2 \neq | f_{i_1,j_1}|^2$.
In this last case, the conditions imposed 
on $B$ by requiring Eq. (\ref{canc2}) to be valid are equivalent to the ones obtained for two distinct jump operators.
So we can reproduce the reasoning done at the beginning of this section and state that {\bf weak irreducibility}
implies {\bf irreducibility} also in the case of Lindblad operators composed by two jumps between
energy eigenstates, apart from cases in which particular criteria for the values of the couplings $f_{l,l'}$
are met.
The calculations for the case in which some of the Lindblad operators are the linear combination of more than two 
jumps allows to derive conditions on the $f_{l,l'}$ linking the matrix elements of $B$ associated to 
transitions between the levels connected by the jumps, similarly to what observed
in the two jumps case.
To summarize, when the master equation is an effective description of the dynamics induced by the weak coupling of the system 
with a thermal bath, the generator belongs to the adjoint-stable subset of the ergodic maps and tipically satisfies irreducibility.
Hence not only the associated restrictions  ${\cal G}$~(\ref{oldGAMMA}) of a thermalizing generator  is invertible on $\mathfrak{L}^{0}_{\text{S}}$, but 
thanks to Eq.~(\ref{SUMRULE}) this property is also shared by all the sums of an arbitrary collection of thermalizing generators.

 \subsection{Derivation of ~Eqs.~(\ref{eq:dyn2}) }\label{DERIVATIONAPPA} 
 Equipped with the results derived in the previous subsection it is now easy to 
 explicitly show how to reformulate the master equation~(\ref{eq:master_eq_rho}) in the form~(\ref{eq:dyn2})
 with the super-operator ${\cal G}_{\vec{u}(t)}$ being invertible on their domain of definition.

 As anticipated  in the main text, we can obtain this by imposing that for all choices of  the control
 vectors $\vec{u}\in  \mathbb{D}$ for which ${\cal D}_{\alpha,\vec{u}}$ is not explicitly null, we ask 
 those super-operators to be thermalizing with fixed point provided by 
  \begin{eqnarray} \label{GIBBSalpha} 
\rho^{\text{(eq)}}_{\alpha;\vec{u}} \equiv \exp[ -{\beta_\alpha} H_{\vec{u}(t)}]/Z_{\alpha;\vec{u}} \;, \qquad Z_{\alpha;\vec{u}} \equiv \mbox{Tr}[\exp[ -{\beta_\alpha} H_{\vec{u}}]]\;.
\end{eqnarray} 
 From the physical point of view this is a rather natural requirement to ask: it simply tell us that,
 by putting S in contact with bath $\alpha$, the model
 will reach the steady state configuration defined by the corresponding thermal state
 $\rho^{\text{(eq)}}_{\alpha;\vec{u}}$.
 As discussed in the previous section, in this setting it is natural to consider the dissipator  ${\cal D}_{\alpha,\vec{u}}$ 
 to be adjoint stable and irreducible.
From~(\ref{SUMRULE}) it  follows that for all assigned $\vec{u}(t)$, also the super-operator $\mathcal{L}_{\vec{u}(t)}$ 
is irreducible and adjoint-stable, due to the fact that according to Eq.~(\ref{eq:master_eq_rho}) it is given by the sum of the ${\cal D}_{\alpha,\vec{u}(t)}$'s plus an irrelevant  Hamiltonian contribution which
 plays no role in deciding these properties, i.e.
\begin{equation}
{\cal L}_{\vec{u}(t)} = 
     {\cal H}_{\vec{u}(t)} 
  + \sum_{\alpha=1}^N \mathcal{D}_{\alpha, \vec{u}(t)}    \label{eq:master_eq_rho111new} \;,
\end{equation}
where we used ${\cal H}_{\vec{u}(t)}$ to identify the commutator with $H_{\vec{u}(t)}$, i.e. ${\cal H}_{\vec{u}(t)}[\cdots]\equiv -i [H_{\vec{u}(t)}, \cdots ]$.
 Therefore, introducing $\rho^\text{(eq)}_{\vec{u}(t)}$ as the unique fixed point of 
$\mathcal{L}_{\vec{u}(t)}$ and invoking~(\ref{IDENTITY1}),
 we can again write 
\begin{eqnarray} \label{IDENTITY1alphaL} 
{\cal L}_{\vec{u}(t)} [\rho] = {\cal G}_{\vec{u}(t)} [\tilde{\rho} -\tilde{\rho}^{(eq)}_{\vec{u}(t)} ]\;,
\end{eqnarray} 
for all $\rho$, with 
 \begin{eqnarray}{\cal G}_{\vec{u}(t)} \equiv - \mathcal{Q}\circ {\cal L}_{\vec{u}(t)}  \circ \mathcal{Q} =-{\cal L}_{\vec{u}(t)}  \circ \mathcal{Q}\;,\label{oldGAMMAalphaL}
\end{eqnarray} 
which is also invertible on the traceless operator set $\mathfrak{L}^{0}_{\text{S}}$.
Specified in the case where $\rho$ is the evolved density matrix of S at time $t$ during its interaction with the baths of the model, and using the property $\partial_t \rho(t) = \partial_t \tilde{\rho}(t)$, we can finally rewrite (\ref{eq:master_eq_rho})
in the form (\ref{eq:dyn2}).

\section{Asymptotic solutions of the fast periodically driven ME}
\label{bla}

Here we discuss the asymptotic solutions of the system ME. 
We start in Sec.~\ref{PERIOD} formally introducing the notion of   limit cycle  solutions, valid for arbitrary driving speed.
Then in Sec.~\ref{appfast} we give a formal derivation of Eq.~(\ref{eq:p0}) of the main text, valid in the fast-driving regime. 
Finally in Sec.~\ref{newappa} we show that any sub-protocol extract from a  fast cyclic control also fulfils the fast limit condition.

\subsection{Periodic driving} \label{PERIOD} 
 
 Consider the case where the control vector $\vec{u}(t)$ of our model, and hence
 the generator  ${\cal L}_{\vec{u}(t)}$ of Eq.~(\ref{eq:master_eq_rho}), 
 is periodic with 
 period $T$, i.e. $ 
 \vec{u}(t+T) = \vec{u}(t)$, for all $t$.
 We have already commented in Appendix~\ref{app:projection}, that requiring 
the ${\cal D}_{\alpha,\vec{u}(t)}$s to be irreducible and adjoint stable, ensures that  ${\cal L}_{\vec{u}(t)}$ is
irreducible and adjoint-stable.
Invoking Theorem 2 of  Ref.~\cite{Menczel2019} we can thus claim that our ME~
admits a limit cycle  solution  ${\rho}_{\vec{u}(t)}^{\text{(lc)}}= 
{\rho}_{\vec{u}(t+T)}^{\text{(lc)}}\in \mathfrak{S}_S$  that is independent from the initial conditions of S, and such that 
\begin{eqnarray} \label{limitcycl} 
\lim_{t\rightarrow \infty}  (\rho(t) - {\rho}_{\vec{u}(t)}^{\text{(lc)}}) =0\;, 
\end{eqnarray} 
the convergency being evaluated e.g.  in  the trace norm.

 \subsection{Fast driving limit}  \label{appfast} 
 In the fast cyclic driving  limit we assume that
 the period $T$ of the cyclic driving $\vec{u}(t)$ is the shortest timescale appearing in the master equation. In this scenario we want to show that, up to linear correction in $T$, 
 we can approximate the  limit cycle solution 
$\rho_{[\vec{u}]}^{\text{(lc)}}(t)$  with a term which is constant in time. 

To begin with, we decompose $\rho(t) = \rho_\text{d}(t) + \rho_\text{nd}(t)$, where $\rho_\text{d}(t)$ denotes the diagonal part of $\rho(t)$ in the energy eigenbasis of $H_{\vec{u}}(t)$, and $\rho_\text{nd}(t)$ describes the non-diagonal terms. Since we assume that $[H_{\vec{u}_1},H_{\vec{u}_2}]=0$, the eigenbasis is constant in time and the dynamics of $\rho_\text{d}(t)$ and $\rho_\text{nd}(t)$ decouple \cite{Breuer2002}.
The heat currents, defined in Eq.~(\ref{eq:heat_rho}), only depend on $\rho_\text{d}(t)$, so we can effectively neglect $\rho_\text{nd}(t)$ and restrict our analysis to $\rho_\text{d}(t)$. We therefore define the following timescale
\begin{eqnarray}\label{DEFTAUE} 
\eta_{[\vec{u}]} \equiv \max_{t\in I_{[\vec{u}]}} \| {\cal G}_{\vec{u}(t)}\| \;,
\end{eqnarray}
where the maximum is taken on period and
where  $\| \cdots \|$ is a norm defined as
\begin{eqnarray}
\| {\cal G}_{\vec{u}(t)}\|\equiv \max_{\Theta \in \mathfrak{L}_{\text S,d}}
\frac{\|{\cal G}_{\vec{u}(t)}[\Theta]\|_1 }{\|\Theta \|_1} \;,
\end{eqnarray} 
where $\| \Theta \|_1\equiv \sqrt{\mbox{Tr}[ \Theta^\dag \Theta]}$ and where $\mathfrak{L}_{\text S,d}$ is the vector space of \textit{diagonal} linear operators. Since the dynamics of $\rho_\text{d}(t)$ only depends on the dissipators, also $\eta_{[\vec{u}]}$ is solely determined by the dissipators. Physically, it thus represents the rate of the fastest possible relaxation to steady state.
%\textcolor{red}{ To begin with, let us observe that according to Eq.~(\ref{eq:dyn2}), we can identify the
%times scale of the system dynamics associated with the driving of the model by using the
%  quantity
%\begin{eqnarray}\label{DEFTAUE} 
%\eta_{[\vec{u}]} \equiv \max_{t\in I_{[\vec{u}]}} \| {\cal G}_{\vec{u}(t)}\| \;,
%\end{eqnarray}
%where the maximum is taken on period and
%where  $\| \cdots \|$ is the induced trace-norm for super-operators~\cite{Watrous2018}, i.e.
%\begin{eqnarray}
%\| {\cal G}_{\vec{u}(t)}\|\equiv \max_{\Theta \in \mathfrak{L}_{\text S}}
%\frac{\|{\cal G}_{\vec{u}(t)}[\Theta]\|_1 }{\|\Theta \|_1} \;,
%\end{eqnarray} 
%with $\| \Theta \|_1\equiv \sqrt{\mbox{Tr}[ \Theta^\dag \Theta]}$. Since the system is finite dimensional, any other choice will be fine as well, the above choice however allows for some simplification. In particular,} 
By direct integration of Eq.~(\ref{eq:dyn2}) on a generic interval 
$[t_1,t_2]$ we can write 
\begin{eqnarray}\label{INEq1} 
\| \rho(t_2) -\rho(t_1) \|_1 = \| \int_{t_1}^{t_2} dt' {\cal G}_{\vec{u}(t')}
\left[\tilde{\rho}^\text{(eq)}_{\vec{u}(t')}- \tilde{\rho}(t') \right] \|_1 \leq 
 \int_{t_1}^{t_2} dt' \| {\cal G}_{\vec{u}(t')}\|
\| \tilde{\rho}^\text{(eq)}_{\vec{u}(t')}- \tilde{\rho}(t')  \|_1 \leq (t_2-t_1) \; \eta_{[\vec{u}]}\;,
\end{eqnarray} 
which explicitly shows that the speed of variation of the density matrix of S along the trajectory
induced by control $\vec{u}(t)$ is explicitly upper bounded by $\eta_{[\vec{u}]}$. 
Accordingly we now formally identify the fast-driving regime by restricting the analysis
to those protocols which fulfil the constraint 
\begin{eqnarray}  \label{fastregime} 
\eta_{[\vec{u}]} T \ll 1\;. 
\end{eqnarray} 
Next we prove that, in the cyclic and fast driving regime, $\rho_{[\vec{u}]}^{\text{(lc)}}(t)$ is approximately given by ${\rho}_{[\vec{u}]}^{\text{(0)}}$.
Accordingly, we can use
 Eq.~(\ref{INEq1}) 
to claim that the distance between ${\rho}_{[\vec{u}]}^{\text{(lc)}}(t)$ and ${\rho}_{[\vec{u}]}^{\text{(lc)}}(t^*)$ is upper bounded by  $|t-t^*| \; \eta_{[\vec{u}]}$. 
Taking a fixed value of $t^*$ and an arbitrary time $t \in [t^*, t^* + T]$ we have that 
\begin{eqnarray} \label{brutta} 
\| {\rho}_{[\vec{u}]}^{\text{(lc)}}(t) - {\rho}_{[\vec{u}]}^{\text{(lc)}}(t^*) \|_1 \leq 
 \eta_{[\vec{u}]} {T} \;.
\end{eqnarray}  
Therefore, invoking the fast driving condition in Eq.~(\ref{fastregime}), we find that at zeroth order in $ \eta_{[\vec{u}]}T$, all values of ${\rho}_{[\vec{u}]}^{\text{(lc)}}(t)$ are given by the same fixed state, which we denote with $\rho_{[\vec{u}]}^{(0)}$.
We now want to show that, up to first order corrections in $ \eta_{[\vec{u}]}T$, the constant term $\rho_{[\vec{u}]}^{(0)}$ is given by Eq.~(\ref{eq:p0}) of the main text.  For this purpose 
 let notice that 
since ${\cal L}_{[\vec{u}]}  \equiv \int_{I_{[\vec{u}]}} {\cal L}_{\vec{u}(t)} dt$ is
a positive sum of ${\cal L}_{\vec{u}(t)}$ which in our construction are irreducible and 
adjoint-stable. Fuethermore, from 
  (\ref{SUMRULE}) we can also 
claim that such superar-operator fulfils the same property. Accordingly 
\begin{eqnarray} 
 {\cal G}_{[\vec{u}]} \equiv - {\cal Q} \circ  {\cal L}_{[\vec{u}]}  \circ {\cal Q}
=
 - {\cal Q} \circ \int_{I_{[\vec{u}]}} {\cal L}_{\vec{u}(t)} dt \circ {\cal Q}=\int_{I_{[\vec{u}]}}  {\cal G}_{\vec{u}(t)} dt \;,
\end{eqnarray} 
must be invertible on $\mathfrak{L}^{0}_{\text{S}}$. 
Integrating hence~(\ref{eq:dyn2}) 
over the interval ${I_{[\vec{u}]}}$ and considering the limiting cycle solution $\rho_{[\vec{u}]}^{\text{(lc)}}(t)$,
 we get 
 \begin{align}
  \int_{I_{[\vec{u}]}}  {\cal G}_{\vec{u}(t)} \left[\tilde{\rho}^\text{(eq)}_{\vec{u}(t)}- \tilde{\rho}_{[\vec{u}]}^{\text{(lc)}}(t) \right]dt =
  \int_{I_{[\vec{u}]}}   {\partial_t}\tilde{\rho}_{[\vec{u}]}^{\text{(lc)}}(t)dt = \tilde{\rho}_{[\vec{u}]}^{\text{(lc)}}(T)-\tilde{\rho}_{[\vec{u}]}^{\text{(lc)}}(0)=0
  \;, 
 \label{eq:dyn2appr}  
\end{align}
where
the last identity follows from the periodicity of $\rho_{[\vec{u}]}^{\text{(lc)}}(t)$. We therefore have that
 \begin{align} 
\int_{I_{[\vec{u}]}} 
{\cal G}_{\vec{u}(t)} \left[ \tilde{\rho}_{[\vec{u}]}^{\text{(lc)}}(t) \right]dt =  \int_{I_{[\vec{u}]}} 
{\cal G}_{\vec{u}(t)}\left[\tilde{\rho}^\text{(eq)}_{\vec{u}(t)}\right] dt \;. 
 \label{eq:dyn2appr1}  
\end{align}
In the  fast driving limit~(\ref{fastregime}), 
the left-hand-side of the above expression can be approximated as ${\cal G}_{[\vec{u}]}[ \tilde{\rho}^{(0)}_{[\vec{u}]}]$ up to linear correction in $\eta_{[\vec{u}]} {T}$. Accordingly 
in this regime (\ref{eq:dyn2appr1}) allows us to finally write 
\begin{align}
{\cal G}_{[\vec{u}]}\left[ \tilde{\rho}^{(0)}_{[\vec{u}]}\right] \simeq 
\int_{I_{[\vec{u}]}} 
{\cal G}_{\vec{u}(t)}\left[\tilde{\rho}^\text{(eq)}_{\vec{u}(t)}\right] dt   \qquad \Longrightarrow \qquad
 \tilde{\rho}^{(0)}_{[\vec{u}]}\simeq 
{\cal G}_{[\vec{u}]}^{-1} \left[ \int_{I_{[\vec{u}]}} 
{\cal G}_{\vec{u}(t)}\left[\tilde{\rho}^\text{(eq)}_{\vec{u}(t)}\right] dt \right] \;,
 \label{eq:dyn2appr2}  
\end{align}
where we used the above-mentioned  invertibility of ${\cal G}_{[\vec{u}]}$. 
This expression, valid at leading order in the expansion in $ \eta_{[\vec{u}]} T$, corresponds to Eq.~(\ref{eq:p0}). 

\subsection{Sub-protocols of fast driving controls} \label{newappa} 
Here we show that a generic  sub-protocol $\vec{u}_A(t)$ extracted from a cyclic trajectory $\vec{u}(t)$ fulfilling  the fast driving limit condition~(\ref{fastregime}), also fulfils the same condition. 

As detailed in Sec.~\ref{Oosq} of the main text, a generic sub-protocol  $\vec{u}_A(t)$ is constructed from  reduction of $\vec{u}(t)$ on a  proper subset $I_A$ of the fundamental period $I_{[\vec{u}]}$. 
Accordingly, from Eq.~(\ref{DEFTAUE}) it follows that 
\begin{eqnarray}\label{DEFTAUE1} 
\eta_{[\vec{u}]} = \max_{t\in I_{[\vec{u}]}} \| {\cal G}_{\vec{u}(t)}\| \geq  
 \max_{t\in I_{A}} \| {\cal G}_{\vec{u}(t)}\| =  \max_{t\in I_{[\vec{u}_A]}} \| {\cal G}_{\vec{u}_A(t)}\| =
 \eta_{[\vec{u}_A]} \;. 
\end{eqnarray} 
Also, indicating with $T_A$ the measure of the interval $I_A$, we have by construction $T_A\leq T$: 
putting this together the thesis finally follows  via the inequality 
\begin{eqnarray} 
 \eta_{[\vec{u}_A]} T_A \leq \eta_{[\vec{u}]} T \ll 1\;.
\end{eqnarray}

\section{Selection of the infinitesimal protocol}
\label{app:infinitesimal}

As discussed in Sec.~\ref{sec:average_partition}, proving the possibility of fulfilling the 
 condition~(\ref{STRONG}) of the main text, is equivalent to showing that 
starting from a generic curve in $\mathbb R^{D}$ ($D=d-1$) that has a null center of mass, we can always split it into two (non-trivial) sub-curves  such that these still have a null center of mass. 
This result is proven explicitly in Sec.~\ref{main}. Then in Sec.~\ref{app:max_quench} we give a characterization of the maximum number $L$ of time intervals entering in Eq.~(\ref{eq:gap_otto}).

\subsection{Main result} \label{main} 
Let $\gamma \equiv \{  \vec{v}(t) | t \in [0,T] \}$ be a piecewise $C^1$ curve
generated by the function
 $\vec v(t): [0,T] \to \mathbb R^D$,
 that satisfies 
\begin{equation}
    \int_0^T \vec v(t) dt = 0. \label{AppC1}
\end{equation}
We want to show that there exist $k \le D+1$ points on the curve $\vec v(t_1),{\cdots},\vec v(t_k)$ and parameters $\tau_1,{\cdots}, \tau_k>0$ such that 
\begin{equation} \label{thm}
    \int_{t_1}^{t_1 + \tau_1} \vec v(t) dt  + \cdots + \int_{t_k}^{t_k + \tau_k} \vec v(t) dt = 0.
\end{equation}
Indeed, in such case the sub-curve identified by the restriction of $\vec{v}(t)$ to  $[t_1, t_1+\tau_1]\cup \dots \cup [t_k, t_k+\tau_k] $  would have a null center of mass.

We can suppose that this curve does not lay on any hyperplane $V$ strictly contained in $\mathbb R^D$, otherwise we can simply repeat the proof in the smaller space $V \cong \mathbb R^{D-1}$. 

We first notice that, calling $C$ the convex hull of the range of the curve $\{\vec v(t): t \in [0,T]\}$ and $\mathring C$ its interior, then $0 \in \mathring C$. 
Indeed, if by contradiction this is not the case, by the Hahn-Banach theorem \cite{Rudin1964} there exists a unitary vector 
$\vec w \in \mathbb R^D$ such that $\vec w\cdot \vec x \ge 0$ for every $\vec x \in C$. Since $\vec v(t) \in C$ for every $t$, 
 $\vec v(t) \cdot \vec w$ is a non negative function, but using Eq. (\ref{AppC1}) we have
$$   \int\vec v(t) \cdot \vec w dt = \left(\int\vec v(t) dt\right) \cdot \vec w = 0, $$
hence $\vec v(t) \cdot \vec w= 0$  for every $t$.
This is equivalent to saying that $\vec v$ lies on the hyperplane $\{ \vec w\cdot \vec x = 0\}$, that is a contradiction.

Since $0 \in \mathring C,$ we can find $D+1$ points $\vec p_1, {\cdots}, \vec p_{D+1} \in \mathring C$,
\begin{align*}
    \vec p_1 =&~ \varepsilon \begin{pmatrix} 1-\frac 1{D+1}, -\frac 1{D+1}, {\cdots}, -\frac 1{D+1} \end{pmatrix}\;, \\
    \vec p_2 =&~ \varepsilon \begin{pmatrix} -\frac 1{D+1}, 1-\frac 1{D+1}, {\cdots}, -\frac 1{D+1} \end{pmatrix}\;, \\
    \vec p_D =&~ \varepsilon \begin{pmatrix} -\frac 1{D+1}, -\frac 1{D+1}, {\cdots}, 1-\frac 1{D+1} \end{pmatrix}\;,\\
    \vec p_{D+1} =&~ \varepsilon \begin{pmatrix} -\frac 1{D+1}, -\frac 1{D+1}, {\cdots}, -\frac 1{D+1} 
    \end{pmatrix}\;,
\end{align*}
with $\varepsilon$ a positive constant,
so that $\vec p_1,{\cdots},\vec p_{D+1}$ do not lie in any hyperplane and 
\begin{equation} \label{CC1}
    \frac{\vec p_1+{\cdots}+\vec p_{D+1}}{D+1}=0.
\end{equation}
Moreover, since $\vec p_j \in C$, there exist $t_1^j, {\cdots}, t_{m_j}^j \in [0,T]$ and a set of coefficients
$\alpha_1^j, {\cdots}, \alpha_{m_j}^j \in (0,1)$ with $m_j \le D+1, \alpha_1^j+ {\cdots} + \alpha_{m_j}^j = 1$, such that
\begin{equation} \label{CC2}
    \alpha_1^j \vec v(t_1^j) + {\cdots} + \alpha_{m_j}^j \vec v(t_{m_j}^j) = \vec p_j.
\end{equation}
Since the $\vec p_j$ by construction do not lie on any hyperplane, the same holds true for the family $\vec v(t_k^j)$. 
To simplify the notation, we reindex $t_k^j = t_l$, $1\le l\le m \le D+1$, where $m = \sum_{j=1}^{D+1} m_j$, and from \eqref{CC1} and \eqref{CC2}, we notice that there exist positive numbers $a_1, {\cdots}, a_m$ with $a_1 + {\cdots} + a_m =1$ such that 
\begin{equation} \label{CC3}
    a_1\vec v(t_1) + {\cdots} + a_m\vec v(t_m) = 0.
\end{equation}
Moreover, the vectors $\vec v(t_1), {\cdots}, \vec v(t_m)$ do not lie on any hyperplane. Up to reordering, we can assume that $\vec v(t_{m-D+1}), {\cdots},\vec v(t_m)$ are linearly independent. We define the maps 
\begin{equation}
    f_l(\tau) = 
    \begin{cases}
    \int_{t_l}^{t_l+\tau} \vec v(t) dt & \text{ if } t_l < T,\\
    \int_{T-\tau}^{T} \vec v(t) dt & \text{ if } t_l = T,
    \end{cases}
\end{equation}
and
\begin{equation}
    F(\tau_1, {\cdots}, \tau_m) = f_1(\tau_1) + {\cdots} + f_m(\tau_m).
\end{equation}
Then \eqref{thm} is proven if we have that for some choice of $\tau_1, {\cdots}, \tau_m > 0$ arbitrarily small, $F(\tau_1, {\cdots}, \tau_m) =0$. For this purpose, we notice that $\partial_{\tau} f_l(0) = \vec v(t_l)$, and
\begin{equation}
    \nabla_{\tau_{m-D+1},{\cdots},\tau_D} F|_{\tau_{m-D+1}=0, {\cdots}, \tau_m = 0} = \begin{pmatrix} \vec v (t_{m-D+1}), & \cdots & \vec{v}(t_m)   \end{pmatrix}
\end{equation}

which is an invertible matrix, since the vectors $\vec v (s_{m-D+1}), \cdots, \vec{v}(s_m)$ are linearly independent. Moreover, $F(0,{\cdots},0)=0$. Therefore, by the implicit function theorem \cite{Rudin1964}, in a neighborhood of $0$, there exist $C^1$ functions $\sigma_{m-D+1}(\tau_1, {\cdots}, \tau_{m-D}), {\cdots}, \sigma_{D}((\tau_1, {\cdots}, \tau_{m-D}))$ such that 
\begin{equation} \label{Fsigma}
    F(\tau_1, {\cdots}, \tau_{m-D},\sigma_{m-D+1}(\tau_1, {\cdots}, \tau_{m-D}), {\cdots}, \sigma_{D}((\tau_1, {\cdots}, \tau_{m-D}))) = 0
\end{equation}
and $\sigma_j(0,{\cdots},0) =0$. If we have that for an appropriate choice of $\tau_1,{\cdots}, \tau_{m-D} > 0$, then $\sigma_j(\tau_1, {\cdots}, \tau_{m-D}) > 0$ for every $j$, we obtain \eqref{thm}. Differentiating 
$$0 = F(a_1 t, {\cdots}, a_{m-D} t, \sigma_{m-D+1}(a_1 t, {\cdots}, a_{m-D} t), {\cdots}, \sigma_m(a_1 t, {\cdots}, a_{m-D} t))$$ in $t=0$, we obtain 
\begin{equation}
    a_1 \vec v(t_1) + {\cdots} + a_{m-D} \vec v(t_{m-D}) + \frac d {d t} \sigma_{m-D+1} (0) \vec v(t_{m-D+1}) + {\cdots} + \frac d {d t} \sigma_{m} (0) \vec v(t_{m}) = 0.
\end{equation}
Combining this with \eqref{CC3} and recalling that $\vec v(t_{m-D+1}), {\cdots},\vec v(t_m)$ are linearly independent, we obtain that 
\begin{equation}
    \frac d {d t} \sigma_{j} (0) = a_j > 0.
\end{equation}
Therefore, for $\delta$ small enough, $\sigma_j(\delta a_1, {\cdots}, \delta a_{m-D}) > 0$, and we obtain \eqref{thm} by choosing $\tau_j = a_j \delta$ for $j \le m-D$, and 
$\tau_{m-D+j} = \sigma_{m-D+j}(\delta a_1, {\cdots}, \delta a_{m-D})$, and the analogous choice if $t_j = T$ for some $j$.

\subsection{Maximum number of sudden quenches}
\label{app:max_quench}

As discussed in Sec.~\ref{sec:average_partition} and in App. \ref{app:infinitesimal},
the most powerful protocol is a generalized Otto cycle composed by a finite number $k$ of
infinitesimal segments where the control is constant.
Here we will prove that $k \leq L= D-1$, where $D$ is the dimension of the
vector space in which the function $\vec{v}(t)$ introduced in Sec. \ref{sec:average_partition} lives.

As discussed in the main text, by repeating many times the cut-and-choose procedure we eventually end up with an infinitesimal protocol to which we can associate a curve composed by $k$ infinitesimal segments each one
starting from a vector $\vec{w}_i \in \mathbb{R}^D$ with $0\leq i \leq k$.
We assume this curve to have null center of mass, i.e. that  Eq. (\ref{thm}) is valid. For an infinitesimal protocol, Eq. (\ref{thm}) reduces to
\begin{equation}
    \sum_{i=1}^k \alpha_i\vec{w}_i = 0,
    \label{eq:zer_app}
\end{equation}
where $\alpha_i = \tau_i/T > 0$ is such that $\sum \alpha_i = 1$.

Let $C$ be the convex hull generated by the vectors $\{ \vec{w}_i : i =1,..k\}$. As argued in the previous appendix, Eq.~(\ref{eq:zer_app}) implies 
that $0$ belongs to $\mathring{C}$. 
Assuming that the original vectors do not lie on any hyperplane, by linear independence we can identify a 
subset $S$ of indices of $\{1,2,...,k\}$ and some coefficients $\xi_i $ such that $ \sum_{i \in S} \xi_i\vec{w}_i = 0 $ and $\sum_{i \in S} \xi_i = 1$.
Moreover, $S$ has at most $D+1$ elements.

We now take a linear combination of $ \sum_{i \in S} \xi_i\vec{w}_i = 0 $ and Eq.~(\ref{eq:zer_app}) obtaining, up to reordering the vectors
\begin{equation}
    (\alpha_1 - c\xi_1)\vec{w}_1 + \cdots + (\alpha_{D+1} - c\xi_{D+1})\vec{w}_{D+1} + a_{D+2}\vec{w}_{D+2} + \cdots + a_k\vec{w}_k = 0,
\end{equation}
where $c$ is a positive constant. If $c=0$, all coefficients are positive, whereas for large values of $c$, the first $D+1$ coefficients become negative. We therefore choose the smallest value of $c$ such that one coefficient is null, and the other ones are positive,
that is 
\begin{equation}
    c = \min_{i=1,\cdots D+1} \frac{\alpha_i}{\xi_i}.
    \label{eq:c}
\end{equation}
The following argument can be simply generalized to the case where more than one coefficient is null. We can assume that the index that minimizes the right hand side of Eq.~(\ref{eq:c}) is $i=1$. With this choice, we have that
\begin{equation}
\begin{cases}
    &c\xi_1 \vec{w}_1 + \cdots c\xi_{D+1}\vec{w}_{D+1}  = 0, \\
    &(\alpha_2 - c\xi_2)\vec{w}_2 + \cdots (\alpha_{D+1}-c\xi_{D+1})\vec{w}_{D+1} + \alpha_{D+2}\vec{w}_{D+2} +\cdots + \alpha_{k} \vec{w}_k  = 0. 
\end{cases} \label{2cases}
\end{equation}
We now define two new sub-curves, A and B, of the original infinitesimal curve through Eq.~(\ref{2cases}). The first is given by $D+1$ infinitesimal segments, centered around $\vec{w}_{1}, \dots, \vec{w}_{D+1}$, with time duration $\tau^\text{(A)}_i/T$ given by the coefficients of the first row of Eq.~(\ref{2cases}). The second is centered around $\vec{w}_2,\dots, \vec{w}_k$ with time duration $\tau^\text{(B)}_i/T$ given by the coefficients of the second row of Eq.~(\ref{2cases}). Thanks to Eq.~(\ref{2cases}), these two sub curve have null center of mass [they satisfy Eq.~(\ref{thm})], and thanks to the fact that $\sum_i \tau^\text{(A)}_i + \sum_i \tau^\text{(B)}_i=1$, they are a disjoint partition of the initial curve into two sub-curves.
Each one of the two sub-curves is associated to a sub-protocol, and following the cut-and-choose argument introduced in Sec. \ref{Oosq},
we know that the sub-protocol relative to one or the other sub-curve is not less powerful than the original one. 
If this procedure selects the curve supported on $D+1$ vectors, the proof is over. If the most powerful one is the one composed of $k-1 > D+1$ vectors, we can reiterate this argument until we end up with a protocol made up of $D+1$ vectors.

\section{One control per temperature is sufficient for positive GAPs}
\label{app:positive_GAPs}
Consider the expression of the generalized power \eqref{eq:pow_simple}, which can be rewritten as 

\begin{eqnarray} 
     P_{\bf c}[\{\vec{u}_i,\mu_i\}]&=&\dfrac{\sum_{i,j=1}^L c_{\alpha_i} \pi_i \pi_j 
P_{i\leftarrow j}
}{\sum_{i=1}^L \pi_i}\;, 
\label{eq:pow_simple_app} \\
\end{eqnarray}
where $\alpha_i$ is the constant value of $\alpha(t)$ during the interval $d\tau_i$, and
 $\pi_i\equiv \mu_i  \Gamma_{\vec{u}_i}$.

We assume now that the GAP to maximize is \textit{positive}, that is it consists of a positive average of the currents extracted from some of the thermal sources i.e. $c_{\alpha_i}\geq 0\ \forall i$ (note that this is the case both for the engine and the refrigerator). In such a case it is easy to see that
\begin{equation}
\label{ineq:blockP}
 P_{\bf c}[\{\vec{u}_i,\mu_i\}]\leq \dfrac{\sum_{i,j=1}^L c_{\alpha_i} \pi_i \pi_j \tilde{P}_{i\leftarrow j}}{\sum_{i=1}^L \pi_i}\;, 
\end{equation}
Where $\tilde{P}_{i\leftarrow j}=P_{i\leftarrow j}$ if $\beta_i\neq \beta_j$ while $\tilde{P}_{i\leftarrow j}=0$ if the two temperatures are the same. The inequality holds thanks to the fact that $c_{\alpha_i}\equiv c_{\beta_i}\geq 0$ is positive and only depends on the temperature, plus property~\eqref{eq:Psym_property} (that is, $P_{i\leftarrow j}+P_{j\leftarrow i}\leq 0$ when $\beta_i= \beta_j$). We will now maximize the right hand side of~\eqref{ineq:blockP}, which in the end will result in a maximization of $P_{\bf c}$ as for the optimal control the inequality is saturated.
Consider the list of controls to be ordered in such a way to collect in the first $k$ entries all the points at temperature $\beta_1^{-1}$, i.e.
\begin{equation}
\beta_i=\beta_1 \Leftrightarrow 1\leq i \leq k
\end{equation}
Then the right hand side of~\eqref{ineq:blockP} can be cast as
\begin{equation}
\label{eq:Ptilde1}
\dfrac{\sum_{i\leq k,j>k}\pi_i \pi_j (c_1 \tilde{P}_{i\leftarrow j}+c_j \tilde{P}_{j\leftarrow i})+\sum_{i>k,j>k}c_i\pi_i \pi_j \tilde{P}_{i\leftarrow j}}{\sum_{i\leq k} \pi_i+\sum_{i>k} \pi_i}\ .
\end{equation}
It is possible to re-express the above equation in terms of the time ratios spent on each control point $\mu_i$. Focusing on the controls at temperature $\beta_1^{-1}$ we consider the renormalized fraction of time spent on the first $k$ points, i.e. we introduce 
\begin{equation}
\theta^{(1)}_i=\frac{\mu_i}{\sum_{j=1}^k \mu_j}\ ,
\end{equation}
meaning that the vector $\vec{\theta}^{(1)}$ represents a normalized probability distribution on those points. With this definition we see easily that expression~\eqref{eq:Ptilde1} is in the form
\begin{equation}
\frac{\vec{\theta}^{(1)}\cdot\vec{a}+A}{\vec{\theta}^{(1)}\cdot\vec{b}+B}
\end{equation}
for appropriate definitions of $\vec{a}$, $A$, $\vec{b}$, $B$. When tuning the time fractions spent on each point at temperature $\beta_1^{-1}$ the best option will be thus to concentrate on one point only, i.e. $\theta^{(1)}_i=\delta_{i\bar{i}}$ where $\bar{i}$ reaches the maximum in the following affine inequality (see \textbf{Lemma} below)
\begin{equation}
\frac{\vec{\theta}^{(1)}\cdot\vec{a}+A}{\vec{\theta}^{(1)}\cdot\vec{b}+B}\leq 
\max_i \frac{a_i+A}{b_i+B}\ .
\end{equation}
It is then sufficient to repeat the same argument for each temperature to prove that for any positive GAP at most one control point per temperature is needed in the maximization.

The power can be thus optimized on the form
\begin{equation}
P_{\bf c}[\{\vec{u}_\alpha,\mu_\alpha\}]=\dfrac{\sum_{\alpha,\beta=1}^N c_\alpha \pi_\alpha \pi_\beta 
P_{\alpha\leftarrow \beta}}{\sum_{\alpha=1}^N \pi_\alpha}\;, 
\end{equation}
which is the same as~\eqref{eq:pow_simple}, except for the indices running only on different temperatures. We notice also that if more than one coefficient $c_\alpha$ is zero the optimization can be reduced again. That is, suppose $c_\alpha\neq 0$ for $\alpha=1,\dots,\kappa$ with $\kappa\leq N-2$. Then the above expression takes the form
\begin{equation}
\dfrac{\sum_{\alpha,\beta\leq\kappa} c_\alpha \pi_\alpha \pi_\beta 
P_{\alpha\leftarrow \beta}+\sum_{\alpha\leq\kappa,\beta>\kappa}c_\alpha \pi_\alpha \pi_\beta 
P_{\alpha\leftarrow \beta}}{\sum_{\alpha=1}^\kappa \pi_\alpha+\sum_{\alpha=\kappa+1}^N \pi_\alpha}\;, 
\end{equation}
and the Lemma can be used again to collapse all the last $N-\kappa$ controls in one, remaining with $\kappa+1$ points. This has the immediate consequence, that e.g. for a refrigerator two controls are always sufficient to optimize a refrigerator, where $\kappa=1$.

\paragraph*{\textbf{Lemma}}
Given a probability distribution $\vec{p}$, a vector $\vec{a}$, a positive vector $\vec{b}\geq 0$, a constant $A$ and a positive constant $B\geq 0$, it holds that
\begin{align}
\label{ineq:Lemma}
\dfrac{\vec{p}\cdot \vec{a}+A}{ \vec{p}\cdot\vec{b}+B}\leq \max_i \frac{a_i+A}{b_i+B}\ .
\end{align}

\paragraph*{\textbf{Proof.}} 

First notice that the values of $A$ and $B$ can be reabsorbed in the definitions of $\vec{a}$ and $\vec{b}$, due to $\vec{p}$ being a probability distribution $p_i\geq 0$, $\sum_i^n p_i=1$. Formally
\begin{equation}
a'_i=a_i+\frac{A}{n}\ ,\quad b'_i=b_i+\frac{B}{n} \quad \Rightarrow \dfrac{\vec{p}\cdot \vec{a}+A}{ \vec{p}\cdot\vec{b}+B}=\dfrac{\vec{p}\cdot \vec{a'}}{ \vec{p}\cdot\vec{b'}}
\end{equation}
Now consider a convex combination $\vec{p}=\lambda\vec{p}^{(1)}+(1-\lambda)\vec{p}^{(2)}$ and without loss of generality
\begin{equation}
r^{(1)}\equiv\dfrac{\vec{p}^{(1)}\cdot \vec{a'}}{ \vec{p}^{(1)}\cdot\vec{b'}}\leq \dfrac{\vec{p}^{(2)}\cdot \vec{a'}}{ \vec{p}^{(2)}\cdot\vec{b'}}\equiv r^{(2)}\ .
\end{equation}
It follows that
\begin{align}
\dfrac{\lambda\vec{p}^{(1)}\cdot \vec{a'}+(1-\lambda)\vec{p}^{(2)}\cdot \vec{a'}}{ \lambda\vec{p}^{(1)}\cdot\vec{b'}+(1-\lambda)\vec{p}^{(2)}\cdot\vec{b'}}
=\dfrac{\lambda r^{(1)}\vec{p}^{(1)}\cdot \vec{b'}+(1-\lambda)r^{(2)}\vec{p}^{(2)}\cdot \vec{b'}}{ \lambda\vec{p}^{(1)}\cdot\vec{b'}+(1-\lambda)\vec{p}^{(2)}\cdot\vec{b'}}
\geq \dfrac{\lambda r^{(2)}\vec{p}^{(1)}\cdot \vec{b'}+(1-\lambda)r^{(2)}\vec{p}^{(2)}\cdot \vec{b'}}{ \lambda\vec{p}^{(1)}\cdot\vec{b'}+(1-\lambda)\vec{p}^{(2)}\cdot\vec{b'}}=r^{(2)}\ .
\end{align}
where in the last inequality we used that $\vec{b'}\geq 0$. The same inequality be carried out in the other sense and both inequalities can be condensed as
\begin{align}
\dfrac{\vec{p}^{(1)}\cdot \vec{a'}}{ \vec{p}^{(1)}\cdot\vec{b'}}\
\leq
\dfrac{\lambda\vec{p}^{(1)}\cdot \vec{a'}+(1-\lambda)\vec{p}^{(2)}\cdot \vec{a'}}{ \lambda\vec{p}^{(1)}\cdot\vec{b'}+(1-\lambda)\vec{p}^{(2)}\cdot\vec{b'}}
\leq
\dfrac{\vec{p}^{(2)}\cdot \vec{a'}}{ \vec{p}^{(2)}\cdot\vec{b'}}\ .
\end{align}
If follows in particular that expression is convex and is obtained on the extremal points of the polytope in which $\vec{p}$ lives. The extremal points are deterministic points ($p_i=\delta_{i\bar{i}}$ for some $\bar{i}$), and thus inequality~\eqref{ineq:Lemma} is proven.

\section{Power of many interacting qubits}
\label{app:many_qubits}
In this appendix we prove Eqs.~(\ref{eq:p_asymp}), (\ref{eq:eta_max_pow}) and (\ref{eq:gap_ni_i_fridge}), and the fact that the coefficient of performance of the refrigerator at maximum power is null. We start from the refrigerator case. As discussed in Sec.~\ref{subsec:spectra_opt}, the maximum average cooling power is given by
\begin{equation}
    P_\text{[R]}^\text{(max)} = \frac{1}{\left(\sqrt{\Gamma_{\alpha_1}^{-1}}+\sqrt{\Gamma_{\alpha_2}^{-1}}\right)^2} \frac{1}{\beta_2} \max_{\varepsilon_2} \beta_2\varepsilon_2 \frac{e^{-\beta_2\varepsilon_2}(d-1)}{1+(d-1)e^{-\beta_2\varepsilon_2}}.
    \label{eq:p_fridge_asy_1}
\end{equation}
Defining $x=\beta_2\varepsilon_2$, we need to find the maximum of the function
\begin{equation}
    f(x) = \frac{x e^{-x}(d-1)}{1+(d-1)e^{-x}}.
\end{equation}
Setting to zero the derivative of $f(x)$, the optimal value $x^*$ is determined by solving
\begin{equation}
    e^{x^*} (x^*-1) = d-1.
\end{equation}
The solution to this equation can be given in terms of the Lambert function $W(z)$, which is defined implicitly by the relation $z = W\,e^W$. We therefore find
\begin{equation}
    x^* = 1 + W\left(\frac{d-1}{e}\right).
    \label{eq:x_star}
\end{equation}
Plugging this into $f(x_1)$, and using the relation
$e^{-W(x)} = W(x)/x$,    
we have that
\begin{equation}
    f(x^*) = W\left(\frac{d-1}{e}\right).
    \label{eq:f_star}
\end{equation}
This is an exact solution. The behavior of the Lambert function, for large arguments, is given by
\begin{equation}
    W(z) = \ln(z) - \ln(\ln(z)) + O(1).
\end{equation}
Retaining only the largest contribution in the limit $d\to \infty$, we find
\begin{equation}
    f(x^*) \approx \ln(d).
\end{equation}
Plugging this result into Eq.~(\ref{eq:p_fridge_asy_1}) with $d=2^n$ yields the second relation of Eq.~(\ref{eq:p_asymp}). In the NI case, the cooling power of $n$ qubits is given by $n$ times the power of a single qubit. Setting $d=2$ in Eq.~(\ref{eq:f_star}) and plugging the result into Eq.~(\ref{eq:p_fridge_asy_1}) gives an an exact expression of the maximum cooling power of a single qubit. Equation~(\ref{eq:gap_ni_i_fridge}) is therefore proven by combining this relation with Eq.~(\ref{eq:p_asymp}).

The coefficient of performance $C_\text{op}$ of a refrigerator is defined as the ratio between $P_\text{[R]}^\text{(max)}$, and the power provided to the system. Under the specific protocol considered in Sec.~\ref{subsec:many_body}, it is simply given by
\begin{equation}
    C_\text{op} = \frac{\varepsilon_2}{\varepsilon_1-\varepsilon_2}.
    \label{eq:cop_app}
\end{equation}
The $C_\text{op}$ at maximum power is therefore given by inserting the values of $\varepsilon_1^*$ and $\varepsilon_2^*$ that maximize Eq.~(\ref{eq:p_fridge_asy_1}) into Eq.~(\ref{eq:cop_app}). Since $\varepsilon_1^*\to \infty$ while $\varepsilon_2^*$ is finite, we have that the $C_\text{op}$ at maximum cooling power is null.

We now turn to the heat engine case. As discussed in Sec.~\ref{subsec:spectra_opt}, the maximum average extracted power is given by
\begin{equation}
    P_\text{[E]}^\text{(max)} = \frac{1}{\left(\sqrt{\Gamma_{\alpha_1}^{-1}}+\sqrt{\Gamma_{\alpha_2}^{-1}}\right)^2} \frac{1}{\beta_1\beta_2} \max_{\varepsilon_1 \,\varepsilon_2}
\dfrac{(\beta_1\varepsilon_1\,\beta_2-\beta_2\varepsilon_2\, \beta_1)(e^{-\beta_1\varepsilon_1}-e^{-\beta_2\varepsilon_2})(d-1)}{(1+(d-1)e^{-\beta_1\varepsilon_1})(1+(d-1)e^{-\beta_2\varepsilon_2})}\ .
 \label{eq:p_engine_asy_1}
\end{equation}
Defining $x_1=\beta_1\varepsilon_1$ and $x_2 = \beta_2\varepsilon_2$, we need to find the maximum of the function
\begin{equation}
    f(x_1,x_2) =\frac{(x_1\beta_2-x_2\beta_1)(e^{-x_1}-e^{-x_2})(d-1)}{(1+(d-1)e^{-x_1})(1+(d-1)e^{-x_2})}
\end{equation}
in the limit of $d\to +\infty$. Approximating $d-1 \approx d$ and setting the partial derivatives of $f(x_1,x_2)$ to zero, the optimal values $x_1^*$ and  $x_2^*$ are determined by solving
\begin{equation}
\begin{aligned}
	&\beta_1 e^{x^*_1} (d + e^{x^*_2}) x^*_2  -  \beta_2 \left(e^{2 x^*_1} - d e^{x^*_2} + e^{x^*_1 + x^*_2} ( x^*_1-1) + d e^{x^*_1} (1 + x^*_1)\right)  = 0, \\
	&\beta_2 e^{x^*_2} (d + e^{x^*_1}) x^*_1  -  \beta_1 \left( e^{2 x^*_2} -d e^{x^*_1} + e^{x^*_1 + x^*_2} ( x^*_2-1) + d e^{x^*_2} (1 + x^*_2)\right) = 0.
\end{aligned}
\label{eq:engine_diff}
\end{equation}
We were not able to explicitly find a solution to this set of equation. We therefore search for a perturbative solution, which is valid in the limit of large $d$. We do this by choosing an ``ansatz'' of the form 
\begin{equation}
\begin{aligned}
	x_1^* &= \sum_i a_1^{(i)} g_i(d), \\
	x_2^* &= \sum_i a_2^{(i)} g_i(d),
\end{aligned}
\end{equation}
where $g_i(d)$ are a set of functions that capture the asymptotic behavior of $x_1^*$ and $x_2^*$, and $a^{(i)}_1$ and $a^{(i)}_2$ are a set of coefficients that do not depend on $d$, which we determine by imposing Eq.~(\ref{eq:engine_diff}), i.e. by setting to zero the coefficients of the terms which diverge fastest in $d$. We choose the same set of functions $g_i(d)$ for both $x_1^*$ and $x_2^*$ since this simplifies the calculation, and since it is a reasonable assumption given that $f(x_1,x_2) = f(x_2,x_1)$, provided that we also exchange the temperatures. 

Unfortunately, a straightforward expansion in powers of $d$ does not work. This is due to the fact that we have both polynomial and exponential terms in Eq.~(\ref{eq:engine_diff}). To find a good ``ansatz'', we thus take inspiration from the exact solution $x_1^*$ found in the refrigerator case: indeed, the functions $f(x)$ in the refrigerator case and $f(x_1,x_2)$ in the heat engine case are formally very similar. Using the asymptotic expansion of the Lambert function, from Eq.~(\ref{eq:x_star}) we have that, in the refrigerator case,
\begin{equation}
	x^* \approx \ln{d} -\ln{\ln{d}} + O(1).
\end{equation}
Using this intuition, and performing some attempts, we choose the following ansatz:
\begin{equation}
\begin{aligned}
	x_1^* &= \ln{d} - \ln\ln{d} - \ln{a_1}, \\
	x_2^* &= \ln{d} + \ln\ln{d} + \ln{a_2}.
\end{aligned}
\label{eq:x1_x2_star}
\end{equation}
Plugging this ansatz into Eq.~(\ref{eq:engine_diff}), and retaining only the fastest diverging term with respect to $d$, we find (up to an irrelevant prefactor)
\begin{equation}
\begin{aligned}
	&\left( a_1 \beta_2+ \beta_1 - \beta_2  \right)\, d^2\ln{d}  = 0, \\
	&\left( a_2 \beta_1 + \beta_1 - \beta_2 \right)\,d^2\ln^2{d} = 0.
\end{aligned}
\end{equation}
In order to suppress these fast diverging term, we set the coefficient to zero, finding
\begin{equation}
\begin{aligned}
	a_1 &= \frac{\beta_2 - \beta_1}{\beta_2}, \quad&\quad  a_2 &= \frac{\beta_2 - \beta_1}{\beta_1}.
\end{aligned}
\end{equation}
We now have an approximate expression for $x_1^*$ and $x_2^*$ which is asymptotically correct. Retaining only the fastest diverging term with respect to $d$, we find
\begin{equation}
	f(x_1^*,x_2^*) \approx (\beta_2-\beta_1)\ln{d}.
	\label{eq:fx1x2}
\end{equation}
Using Eq.~(\ref{eq:fx1x2}) to evaluate Eq.~(\ref{eq:p_engine_asy_1}), and setting $d=2^n$ proves the first relations in Eq.~(\ref{eq:p_asymp}).

The efficiency of a heat engine is defined as the ratio between $P_\text{[E]}^\text{(max)}$ and the heat flux provided by the hot bath. 
Under the specific protocol considered in Sec.~\ref{subsec:many_body}, it is simply given by
\begin{equation}
    \eta = 1-\frac{\varepsilon_2}{\varepsilon_1}.
    \label{eq:eta_app}
\end{equation}
The efficiency at maximum power $\eta(P_\text{[E]}^\text{(max)})$, defined as the efficiency while performing the protocol that maximizes the power, is simply given by Eq.~(\ref{eq:eta_app}) computed in the values $\varepsilon_1^*$ and $\varepsilon_2^*$ that maximize Eq.~(\ref{eq:p_engine_asy_1}). It can be expressed in terms of $x_1^*$ and $x_2^*$ as
\begin{equation}
    \eta(P_\text{[E]}^\text{(max)}) = 1-\frac{\beta_1 x_2^*}{\beta_2x_1^*}.
    \label{eq:eta_pmax_app}
\end{equation}
Plugging Eq.~(\ref{eq:x1_x2_star}) into Eq.~(\ref{eq:eta_pmax_app}), using $d=2^n$ and retaining leading order contributions for large $n$ yields Eq.~(\ref{eq:eta_max_pow}).

\section{Qutrit model}
\label{app:qutrit} 
In this appendix we present the model we employ to describe a qutrit in the Markovian regime. The Hamiltonian of the system is given by Eq.~(\ref{eq:h_qutrit}), and the dynamics of the local density matrix is described by Eq.~(\ref{eq:master_eq_rho}). 
Following the standard derivation of the Lindbald master equation \cite{Breuer2002}, and accounting for the fact that we only couple one bath at the time to the qutrit, we write the total dissipator $\mathcal{D}_{\vec{u}(t)}[{{\rho}}] = \sum_\alpha \mathcal{D}_{\alpha,\vec{u}(t)}[{{\rho}}]$ as
\begin{equation}
    \mathcal{D}_{\vec{u}(t)}[{{\rho}}] := \sum_{i\neq j}\Gamma_{ij}(\vec{u}(t),\alpha(t)) \left({A}_{ij}{{\rho}}{A}_{ij}^\dagger -\frac{1}{2}\left[{A}_{ij}^\dagger {A}_{ij}, {{\rho}}(t)  \right]_+\right)\;, 
    \label{eq:qutrit_dissipator}
\end{equation}
where $\alpha(t)$ is an additional control labelling the bath we are coupled to, and $[\cdots, \cdots]_+$ denotes the anti-commutator operations. $\Gamma_{ij}(\vec{u}(t),\alpha(t))$ is the dissipation rate induced by reservoir $\alpha$ which describes a transition from eigenstate $\ket{i}$ to eigenstate $\ket{j}$ of ${{H}}$ and ${A}_{ij} = \ket{i}\bra{j}$.

We now define the occupation probabilities $p_n(t) = \bra{n}{{\rho}}(t) \ket{n}$. By projecting Eq.~(\ref{eq:master_eq_rho}), provided with Eq.~(\ref{eq:qutrit_dissipator}), onto the eigenstates of ${H}$, we can derive a closed set of equations for $p_n(t)$, given by Eq.~(\ref{eq:qutrit_pauli}).
Explicitly, we have that
\begin{equation}
    \begin{pmatrix}
        {\partial_t}{p}_1 \\ {\partial_t}{p}_2 \\ {\partial_t}{p}_3
    \end{pmatrix} 
    =
    \begin{pmatrix}
        -\Gamma_{12} - \Gamma_{13} & \Gamma_{21} & \Gamma_{31} \\
        \Gamma_{12}  & -\Gamma_{21} -\Gamma_{23} & \Gamma_{32} \\
        \Gamma_{13} & \Gamma_{23} & -\Gamma_{32}  -\Gamma_{31}
    \end{pmatrix}
    \cdot
    \begin{pmatrix}
        p_1 \\ p_2 \\ p_3
    \end{pmatrix},
    \label{eq:qutrit_me_explicit}
\end{equation}
where we omitted for simplity the arguments of the probabilities and of the rates. 

In general, the probabilities associated with the limiting cycle can be computed solving Eq.~(\ref{eq:qutrit_pauli}) imposing periodic boundary conditions, i.e. $p_n(0) = p_n(T)$, and imposing that $\sum_n p_n = 1$. The instantaneous heat flux flowing out of all baths $J(t) = \sum_\alpha J_\alpha(t)$ can then be computed as
\begin{equation}
	J(t) = \Tr\left[  {{H}}_{\vec{u}(t)}\mathcal{D}_{\vec{u}(t)}\left[{\rho}(t) \right] \right] = \sum_n \epsilon_n(t) \partial_t p_n(t) = \sum_{m\neq n}  \epsilon_n(t) \left[ -p_n(t)\Gamma_{nm}(\vec{u}(t),\alpha(t)) + p_m(t)\Gamma_{mn}(\vec{u}(t),\alpha(t))  \right]
\end{equation}
where, in the last equality, we used Eq.~(\ref{eq:qutrit_pauli}). The average power delivered by the heat engine is then given 
\begin{equation}
	P_\text{[E]} = \frac{1}{T} \int_0^T J(t) dt.
\end{equation}
This is the procedure used for the numerical calculations at finite period $T$.

In order to simplify this calculation in the fast-driving regime, we first need to cast Eq.~(\ref{eq:qutrit_me_explicit}) into a form equivalent to Eq.~(\ref{eq:dyn2}). We do this using a slightly different approach respect to the one detailed in the main text, that yields equivalent results. Inserting the relation $p_1 = 1 - p_2 - p_3$ into Eq.~(\ref{eq:qutrit_me_explicit}), and using the detailed balance condition (\ref{eq:dbe}), we find
\begin{equation}
	\partial_t \hat{p}(t) = G(\vec{u}(t),\alpha(t)) \cdot(\hat{p}(t)- \hat{p}^\text{(eq)}_{\alpha(t);\vec{u}(t)}),
    \label{eq:qutrit_me_inv}
\end{equation}
where we defined $\hat{p}(t) = (p_2(t),p_3(t))$, and where $\hat{p}^\text{(eq)}_{\alpha(t);\vec{u}(t)} = ([p^\text{(eq)}_{\alpha(t);\vec{u}(t)}]_2, [p^\text{(eq)}_{\alpha(t);\vec{u}(t)}]_3)$ are the Gibbs probabilities of being in state $\ket{2}$ and $\ket{3}$ when in contact with bath $\alpha(t)$. Omitting the explicit argument, the matrix $G(\vec{u}(t),\alpha(t))$ is given by
\begin{equation}
    G = 
    \begin{pmatrix}
      	\Gamma_{12} + \Gamma_{21} + \Gamma_{23} & +\Gamma_{12} - \Gamma_{32} \\
      	 +\Gamma_{13} - \Gamma_{23} & +\Gamma_{13} + \Gamma_{31} + \Gamma_{32}
    \end{pmatrix},
\end{equation}
which can be shown to be strictly positive definite using the detailed balance condition (\ref{eq:dbe}) and assuming that the bath temperatures are finite. Therefore, any sum of $G(\vec{u}(t),\alpha(t))$ at different times will be positive definite, thus invertible. All relations in the main text thus hold by replacing $\rho(t)$ with $p(t)$, $\tilde{\rho}(t)$ with $\hat{p}(t)$ and ${\cal G}_{\vec{u}(t)}$ with $G(\vec{u}(t),\alpha(t))$. Specifically, defining $\hat{\epsilon}(t) = (\epsilon_2(t),\epsilon_3(t))$, Eqs.~(\ref{eq:gap_otto}) and (\ref{eq:p0_otto}),  become in this notation
\begin{equation} 
 P_\text{[E]}[\{\vec{u}_i,\mu_i\}]  =  \sum_{j=1}^L \mu_j \,\, \hat{\epsilon}^T_j \cdot G(\vec{u}_j,\alpha_j)\cdot( \hat{p}_{[\{\vec{u}_i,\mu_i\}]}^{\text{(0)}}-\hat{p}^\text{(eq)}_{\alpha(t);\vec{u}(t)}),
 \label{eq:qutrit_1}
\end{equation}
and
\begin{equation}
   \hat{p}_{[\{\vec{u}_i,\mu_i\}]}^{\text{(0)}} = \left(\sum_{j=1}^L \mu_j{G}(\vec{u}_j, \alpha_j) \right)^{-1}
    \left[ \sum_{j=1}^L \mu_j {G}(\vec{u}_j,\alpha_j)\cdot \hat{p}^\text{(eq)}_{\alpha_j;\vec{u}_j}\right]\;,
\label{eq:qutrit_2}
\end{equation}
where $\hat{\epsilon}_j$ denotes the value of the energies during the $\text{j}^\text{th}$ time interval, as determined by the control $\vec{u}_j$. We can therefore compute the power of a Generalized Otto cycle in the fast driving regime using Eqs.~(\ref{eq:qutrit_1}) and (\ref{eq:qutrit_2}), which is much easier than solving the dynamics explicitly. The optimization of $ P_\text{[E]}[\{\vec{u}_i,\mu_i\}] $ is then performed as described in Sec.~\ref{sec:qutrit}.

\end{widetext}
\end{appendix}


\begin{thebibliography}{}
\bibitem{NOTA}
Property~\eqref{eq:Psym_property} can also be inferred by the fact that if the temperature of the baths at the time intervals $d\tau_j$ and $d\tau_k$ are the same and equal to $1/\beta$, then $P_{j\leftarrow k}+P_{k\leftarrow j}=-\beta^{-1}\left(S(\tilde{\rho}^{\rm (eq)}_{\vec{u}_k}\parallel\tilde{\rho}^{\rm (eq)}_{\vec{u}_j}) + S(\tilde{\rho}^{\rm (eq)}_{\vec{u}_j}\parallel\tilde{\rho}^{\rm (eq)}_{\vec{u}_k})\right)$.
\end{thebibliography}
\end{document}